\documentclass[12pt]{article}
\usepackage{graphicx}
\usepackage{amsmath}
\usepackage{amssymb}
\usepackage{ulem}
\usepackage{enumerate}
\usepackage{cite}
\usepackage{latexsym}
\begin{document}
\begin{center}
{\bf {\large{A Unique Bosonic Symmetry in a  4D Field-Theoretic System}}}

\vskip 3.0cm

{\sf  R. P. Malik$^{(a,b)}$}\\
$^{(a)}$ {\it Physics Department, Institute of Science,}\\
{\it Banaras Hindu University, Varanasi - 221 005, India}\\

\vskip 0.1cm

$^{(b)}$ {\it DST Centre for Interdisciplinary Mathematical Sciences,}\\
{\it Institute of Science, Banaras Hindu University, Varanasi - 221 005, India}\\
{\small {\sf {e-mails: rpmalik1995@gmail.com; malik@bhu.ac.in}}}
\end{center}

\vskip 2.0 cm

\noindent
{\bf Abstract:} For the {\it combined} field-theoretic system of the four (3 + 1)-dimensional (4D) Abelian 3-form and 1-form gauge theories,
we show the existence of a {\it unique} bosonic symmetry transformation that is constructed from the {\it four} infinitesimal,
continuous and off-shell nilpotent symmetry transformations which exist for the Becchi-Rouet-Stora-Tyutin (BRST) quantized
versions of the {\it coupled} (but equivalent) Lagrangian densities that describe our present 4D field-theoretic system.  The above off-shell nilpotent symmetry 
transformations are nothing but the BRST, co-BRST, anti-BRST and anti-co-BRST, under which, the Lagrangian densities transform to the
total spacetime derivatives. The proof of the {\it uniqueness} of the above bosonic symmetry transformation operator
crucially depends on the validity of {\it all} the four Curci-Ferrari (CF) type restrictions that exist on our theory. 
 We highlight the importance of {\it these} CF-type restrictions, at various levels of our
theoretical discussions, in the context of the {\it unique} bosonic symmetry transformation operator. We compare  {\it this} observation against  the
backdrop of the {\it three} CF-type restrictions that appear in
the requirement of the absolute anticommutativity between the {\it specific} set of 
a couple of nilpotent symmetry transformation operators.

\vskip 0.8cm
\noindent
PACS numbers: 11.15.-q; 12.20.-m;
 03.70.+k \\

\vskip 0.5cm
\noindent
{\it {Keywords}}: Combined field-theoretic system of the 4D BRST-quantized free Abelian 3-form and 1-form gauge theories; coupled 
Lagrangian densities; off-shell nilpotent symmetries; a unique bosonic symmetry; ghost-scale symmetry; CF-type restrictions

\newpage

\section {Introduction}

It is an undeniable truth that the ideas of global and local symmetries (see, e.g. [1-3] and references therein) have played a key role in the development of theoretical physics (since the very inception of the subject of physics as  a branch of modern-day science). In particular, the 4D standard model of 
elementary particle physics, where there is a stunning degree of agreement between theory and experiment, owes a great deal to the
{\it interacting} non-Abelian 1-from (i.e. $p = 1$)
gauge  theory which is based on the {\it local} Yang-Mills gauge symmetries. One of the key theoretical methods for the
covariant canonical quantization of such kinds of theories (based on the gauge symmetries) is the Becchi-Rouet-Stora-Tyutin (BRST) formalism [5-7] where the
unitarity and quantum gauge (i.e. BRST) invariance are respected 
{\it together} at any arbitrary order of perturbative computations (see, e.g. [8]).

The central theme of our present endeavor is the four (3 + 1)-dimensional (4D) BRST-quantized version of the {\it combined} field-theoretic system of 
the free Abelian 3-form\footnote{The study of the higher $p$-form ($p = 2, 3... $) gauge theories has become quite important because such kinds 
of {\it basic} fields appear in the quantum excitations of the (super)string theories which are the forefront areas of research activities in 
theoretical high energy physics (see, e.g. [9-11] and references therein).}
and 1-form gauge theories where we focus on the continuous as well as the discrete symmetry transformation operators and their
algebraic structure(s). In particular, we concentrate on (i) the {\it four} continuous off-shell nilpotent symmetry transformation operators, (ii) a couple of 
discrete duality symmetry transformations, (iii) a {\it unique} continuous bosonic symmetry transformation operator (that is constructed from the 
appropriate anticommutator(s) between the above nilpotent symmetry transformation operators), (iv) the infinitesimal and continuous ghost-scale 
symmetry transformation operator, and (v) the derivation of the precise algebraic structures that
are obeyed by the above continuous and discrete duality symmetry transformation operators. We demonstrate that there is an uncanny resemblance 
between the algebraic structures
that are obeyed by (i) the 
discrete and continuous symmetry transformation operators of our 4D BRST-quantized theory, and (ii) the de Rham cohomological operators\footnote{A set of three operators ($d, \; \delta, \;\Delta$), defined on a compact manifold without a boundary, is christened
as the de Rham cohomological operators of differential geometry where $(\delta)d$ (with $ \delta = \pm\, *\; d\; *,\; 
d = \partial_\mu \; (dx^\mu), \; \delta^2 = 0,  \;d^2 = 0$) are  the (co-)exterior derivatives and $\Delta = \{d,\; \delta \}$ stands for the Lalacian operator.
Here the symbol $*$ corresponds to the Hodge duality operator.
These operators obey an algebra: $d^2 = 0, \; \delta^2 = 0,\; \Delta = \{d,\, \delta \}, \; [\Delta, \, d] = 0, \; [\Delta, \, \delta] = 0$ which is popularly
known as the Hodge algebra [12-14].} 
of differential geometry (see, e.g. [12-14]).

The main motivating factors behind the choice of the 4D BRST-quantized field-theoretic system of the {\it combined} free Abelian 3-form and 1-form
gauge theories are as follows. First of all, for the existence of the off-shell nilpotent (anti-)co-BRST symmetry transpositions in our
4D BRST-quantized theory, it is essential to have the {\it sum} of the BRST-quantized versions of the Lagrangian  densities for the free Abelian 3-form
and 1-form gauge theories {\it together} (see, also Sec. 6 for our comments). It turns out that
the sum of the variation of (i) the kinetic term for the Abelian 3-form
gauge field, and (ii)  the FP-ghost term for the Abelian 1-form gauge field, become the total spacetime derivative under the (anti-)co-BRST symmetry
transformations. Second, a close look at the discrete duality symmetry transformations (13) demonstrates that, in our present theory, 
the {\it basic} Abelian 3-form and 1-form gauge fields are {\it dual} to each-other. Third, the numeral factors that appear in the duality
transformations: $ $ owe their mathematical origin to the Hodge duality $*$ operation on the {\it basic} free Abelian 1-form
and 3-form gauge fields, respectively,  in the 4D flat Minkowskian spacetime [see, also, the footnote after equation (16)]. Fourth, 
the discrete and continuous symmetry transformation operators of our {\it present} 4D BRST-quantized system obey a 
set of very beautiful algebraic structures (cf. Secs. 3-5). Finally, if the numbers and varieties of symmetry transformations,
respected by a theory, are the hallmarks of the beauty (of {\it this} theory), we observe that our present 4D BRST-quantized field-theoretic system
belongs to this  specific {\it class} because it respects (i) a set of {\it six} continuous symmetry transformations, and (ii) a couple
of very useful discrete duality symmetry transformations  (besides a couple of bosonic symmetry transformations 
that {\it change} the ghost numbers of the specific fields  of our 4D BRST-quantized theory [cf. Eq. (35) below]).

Our present investigation is essential and important on the following counts. First of all, in our earlier works [15-17], we have been able to show the
existence of a set of {\it four} infinitesimal, continuous and off-shell nilpotent (anti-)BRST and (anti-)co-BRST symmetry transformation operators for our
present 4D BRST-quantized field-theoretic {\it combined} system of the free Abelian 3-form and 1-from gauge theories. We are curious to know the algebraic
structure that is obeyed by these {\it fermionic} symmetry operators in our present endeavor. Second, we have also been able to demonstrate that
our 4D theory respects a couple of discrete duality symmetry transformation operators. Thus, we are very eager to know {\it their} roles in the 
whole gamut of the algebraic structures that are respected by various 
kinds of symmetry transformation operators of our theory. Third, we lay emphasis on the roles of the CF-type
restrictions on our 4D BRST-quantized theory (as far  as the whole beauty of the algebraic structure(s) is concerned). Fourth, out of the {\it six} 
anticommutators
that exist amongst the {\it four} nilpotent symmetry transformation operators, it is very interesting to show that (i) the {\it two} of them absolutely
anticommute with each-other (modulo the U(1) {\it vector} gauge symmetry-type transformations\footnote{We observe that these specific anticommutators
generate the infinitesimal and continuous symmetry transformations that change the ghost numbers of the specific set of fields 
(e.g. $\phi_\mu, \; \widetilde \phi_\mu, \; \beta_\mu, \; \bar\beta_\mu $) of our 4D BRST-quantized field-theoretic system (cf. Sec. 4 below for more details).}), 
(ii) the {\it other} two anticommutators are found to be
{\it zero} provided we use the {\it specific} set of {\it three} (out of the
total {\it four}) CF-type restrictions that exist on our 4D theory, and (iii) the rest of the {\it two} anticommutators define the {\it true}
bosonic symmetry transformation operators. Finally, we establish that, out of the {\it above} two bosonic symmetry  operators, only
{\it one} of them is independent
on the subspace of quantum fields where {\it all} the four CF-type restrictions are satisfied.

The key theoretical contents of our present endeavor 
are organized as follows. In the next section, we recapitulate the bare essentials of the {\it four} 
nilpotent transformations that exist for our 4D
BRST-quantized field-theoretic system [16,17]. In our Sec. 3, we discuss the algebraic structures of the discrete and continuous symmetry transformation operators and define a set of 
{\it two} bosonic symmetry transformation operators from the {\it four} nilpotent  symmetry transformation operators. In particular, we point out
the appearance of a {\it specific} set of {\it three} (out of the total {\it four}) CF-type restrictions
in the context of the requirement of the absolute anticommutativity between (i) the (anti-)BRST
symmetry transformation operators, and (ii) the (anti-)co-BRST symmetry transformation operators. Our Sec. 4 deals with the bosonic symmetry 
transformations  on (i) the basic fields of our theory, and (ii) the coupled Lagrangian densities,
where we show (i) the appearance of the {\it full} set of the {\it four} CF-type restrictions, and (ii) the proof of the {\it uniqueness} of the bosonic symmetry 
transformation operator. The theoretical content of our Sec. 5 is connected with the discussion on the ghost-scale symmetry transformations. 
Finally, in our Sec. 6, we summarize our key observations
and point out their future perspective.

In our Appendix A, for the sake of readers' convenience, we provide a glossary of all the fields 
(e.g. basic 1-form and 3-form gauge fields, Nakanishi-Lautrup auxiliary fields,
fermioic/bosonic (anti-)ghost fields, additional (axial-)vector fields, etc.) of the coupled (but equivalent) Lagrangian densities 
[cf. Eqs. (1),(2),(7),(8)]
with their characteristic properties (e.g. bosonic/fermionic nature, basic/auxiliary nature, ghost number, etc.).\\


\section{Preliminaries: Off-Shell Nilpotent Symmetries}

We begin  with the following (co-)BRST invariant Lagrangian density ${\cal L}_{(B)} = {\cal L}_{(NG)} + {\cal L}_{(FP)} $ that contains the
non-ghost  part ${\cal L}_{(NG)} $ and the FP-ghost  part ${\cal L}_{(FP)} $ as [15-17]
\begin{eqnarray}\label{1}
{\cal L}_{(NG)}  &=& B \, (\partial \cdot A) - \dfrac{1}{2}\, B^2  +  \dfrac{1}{2} \, B_1^2 +\, B_1\, \Big(\dfrac{1}{3!}\varepsilon^{\mu\nu\sigma\rho} \,\partial_\mu A_{\nu\sigma\rho} \Big) + \dfrac{1}{4}\, B_3^2 - 
\dfrac{1}{2}\, B_3 \big (\partial \cdot \widetilde \phi \big ) \nonumber\\
 &-& \,\dfrac{1}{4} \, \big( B_{\mu\nu} \big)^2 + \dfrac{1}{2}\, B_{\mu\nu} \, \Big[\partial_\sigma A^{\sigma\mu\nu}
+ \dfrac{1}{2}\, \big (\partial^\mu \phi^\nu - \partial^\nu \phi^\mu \big) \Big ] - \dfrac{1}{4}\, B_2^2 + 
\dfrac{1}{2}\, B_2 \big (\partial \cdot \phi \big )   \nonumber\\
&+& \dfrac{1}{4}\, \big( {\cal B}_{\mu\nu}  \big)^2 - \dfrac{1}{2} \, {\cal B}_{\mu\nu} \, \Big[ \varepsilon^{\mu\nu\sigma\rho} \,\partial_\sigma A_\rho 
+ \dfrac{1}{2}\, \big (\partial^\mu \widetilde \phi^\nu - \partial^\nu \widetilde \phi^\mu \big) \Big] , 
\end{eqnarray}
\begin{eqnarray}\label{2}
{\cal L}_{(FP)}  &=&  \dfrac{1}{2}\, \Big [\big (\partial_\mu \bar C_{\nu\sigma} +  \partial_\nu \bar C_{\sigma\mu} 
+ \partial_\sigma \bar C_{\mu\nu} \big ) \big (\partial^\mu C^{\nu\sigma} \big ) +
\big (\partial_\mu \bar C^{\mu\nu}  + \partial^\nu \bar C_1 \big ) f_\nu \nonumber\\ 
&-& \big (\partial_\mu  C^{\mu\nu}  + \partial^\nu  C_1 \big ) \bar F_\nu 
 +  \big( \partial \cdot \bar \beta \big) \, B_4
 - \big( \partial \cdot  \beta \big) \, B_5 - B_4 \, B_5 - 2\, \bar F^\mu\, f_\mu \nonumber\\
&-& \big (\partial_\mu \bar \beta_\nu - \partial_\mu \bar \beta_\nu \big ) 
\big (\partial^\mu \beta^\nu \big )  
-\, \partial_\mu \bar C_2\, \partial^\mu C_2  \Big ] + \partial_\mu \bar C\, \partial^\mu C, 
\end{eqnarray}
where the Abelian 1-form ($A^{(1)} = A_\mu\, dx^\mu $) and 3-form [$A^{(3)} = \frac{1}{3!}\,A_{\mu\nu\sigma}\,(d\,x^\mu \wedge d\,x^\nu \wedge d\, x^\sigma)$] define the free {\it basic} gauge fields\footnote{Our 4D flat spacetime manifold is endowed with the metric tensor $\eta_{\mu\nu} $ = diag (+1, -1, -1, -1) so that the dot
product between two non-null vectors $P_\mu$ and $Q_\mu$ is: $(P \cdot Q) = \eta_{\mu\nu} P^\mu Q^\nu = P_0 Q_0 - P_i Q_i $ where the Greek indices $\mu, \nu, \sigma... = 0, 1, 2, 3 $ and
the Latin indices $i, j, k...= 1, 2, 3 $. We choose the 4D Levi-Civita tensor $\varepsilon_{\mu\nu\sigma\rho} $ such that $\varepsilon_{0123}  = + 1 = - \varepsilon^{0123}$  and, when two of them are contracted together, 
the standard relationships: $\varepsilon_{\mu\nu\sigma\rho} \varepsilon^{\mu\nu\sigma\rho} = -\, 4!, \;
\varepsilon_{\mu\nu\sigma\rho} \varepsilon^{\mu\nu\sigma\eta} = -\, 3!\, \delta^\eta_\rho, \; \varepsilon_{\mu\nu\sigma\rho} \varepsilon^{\mu\nu\eta\kappa} 
= -\, 2! (\delta^\eta_\sigma \delta^\kappa_\rho - \delta^\eta_\rho \delta^\kappa_\sigma )$, etc.,  are satisfied.}
$A_\mu$ and $A_{\mu\nu\sigma}$, respectively. In the Lagrangian density ${\cal L}_{(NG)} $, the set of bosonic 
Nakanishi-Lautrup type auxiliary fields
$B, B_1, B_2, B_3, B_{\mu\nu}, {\cal B}_{\mu\nu}$ have been invoked for the linearization purposes\footnote{For the algebraic
convenience, we slightly {\it differ} from our earlier works [15-17] in the mathematical expressions for (i) the gauge-fixing
terms of the gauge field $A_\mu$ and axial-vector field $\widetilde \phi_\mu$,  and (ii) the kinetic term of the Abelian 1-from gauge field $A_{\mu} $
({\it without} changing the {\it actual} values [15-17] of the Nakanishi-Lautrup type auxiliary fields $B = (\partial \cdot A), 
\; B_3  = (\partial \cdot \widetilde \phi)$ and 
${\cal B}_{\mu\nu} = \varepsilon_{\mu\nu\sigma\rho} \partial^\sigma A^\rho + \frac{1}{2}\, (\partial_\mu \widetilde \phi_\nu 
- \partial_\nu \widetilde \phi_\mu)$, respectively, in terms of the basic 
Abelian 1-form vector gauge field $A_\mu$ and  the axial-vector field $\widetilde \phi_\mu$).}. For instance, the 
auxiliary fields $B_1, \, {\cal B}_{\mu\nu} $ have
been utilized in the linearizations of the kinetic terms for the Abelian 3-form and 1-form free gauge fields, respectively. On the other hand, the set of 
auxiliary fields $B, (B_3) B_2, B_{\mu\nu}$ have been used to linearize the gauge-fixing terms for the Abelian gauge field $A_\mu$, the (axial-)vector
fields $(\widetilde\phi_\mu)\phi_\mu$ (which appear in the theory because of the reducibility properties)	and the Abelian 3-form 
gauge field $A_{\mu\nu\sigma}$, respectively. In the Faddeev-Popov (FP) ghost part of the Lagrangian density ${\cal L}_{(FP)} $, we have the {\it basic} 
(anti-)ghost fields: $(\bar C_{\mu\nu})C_{\mu\nu}, \, (\bar \beta_\mu)\beta_\mu, \, (\bar C_2)C_2, (\bar C_1)C_1, \, (\bar C)C $ 
with the ghost numbers: $(-1)+1, (-2)+2, (-3)+3, (-1)+1, (-1)+1 $, respectively.  The fermionic (anti-)ghost
fields $(\bar C)C $ are associated with the Abelian 1-form gauge field and the rest of the above {\it basic} (anti-)ghost fields are connected  to the
Abelian 3-from gauge field $A_{\mu\nu\sigma}$. In the above, all the {\it basic} (anti-)ghost fields are {\it fermionic} in nature 
{\it except} the (anti-)ghost vector
fields $(\bar \beta_\mu)\beta_\mu$ which are {\it bosonic} (i.e. $\beta_\mu^2 \neq 0, \; \bar \beta_\mu^2 \neq 0,\, \beta_\mu \bar \beta_\nu 
=  \bar \beta_\nu \beta_\mu$, etc.) in nature. We also have fermionic auxiliary (anti-)ghost fields $(\bar F_\mu)f_\mu$
with the ghost numbers (-1)+1 and the bosonic set of auxiliary (anti-)ghost fields $(B_5)B_4$ with the ghost numbers (-2)+2, respectively. 
All the basic and auxiliary (anti-)ghost fields are required for the proof of unitarity in the theory (see, e.g. [8] for details). 
It is quite straightforward to check 
that the following infinitesimal, continuous and off-shell nilpotent (i.e. $s_{(d)b}^2 = 0$) (co-)BRST [i.e. (dual-)BRST]
symmetry transformations $s_{(d)b} $, namely;
\begin{eqnarray}\label{3}
&& s_d A_{\mu \nu\sigma} = \varepsilon_{\mu\nu\sigma\rho}\, \partial^\rho \bar C, 
\quad 
s_d A_\mu = \dfrac{1}{2} \,\varepsilon_{\mu\nu\sigma\rho}\,\partial^\nu \bar C^{\sigma\rho}, \;\; 
s_d  \bar C_{\mu\nu}  =  \partial_\mu \bar \beta_\nu - \partial_\nu \bar \beta_\mu, \;\; s_d \bar f_\mu = \partial_\mu B_5, \nonumber\\
&& s_d \bar {\cal B}_{\mu\nu} = \partial_\mu \bar F_\nu - \partial_\nu \bar F_\mu, \quad
s_d \bar \beta_\mu = \partial_\mu \bar C_2, \;\;  s_d  C_1 = +\,B_3 , \quad s_d \beta_\mu = -\, f_\mu,  
\;\; s_d \widetilde \phi_\mu  = +\, \bar F_\mu, \nonumber\\
&&  s_d  C_{\mu\nu} = + \, {\cal B}_{\mu\nu},   \qquad s_d C = -\, B_1, \quad
s_d C_2 = B_4, \qquad s_d \bar C_1 = B_5, \qquad s_d F_\mu = +\, \partial_\mu B_3, \nonumber\\
&&s_d \; \Bigl [ \bar C_2,\, \bar C,\, f_\mu,\, {\bar F}_\mu,\, \phi_\mu, \, B,\, B_1, \, B_2,\, B_3, \,  B_4,\, B_5,\, B_{\mu\nu},  \, \bar B_{\mu\nu},
\, {\cal B}_{\mu\nu}\Bigl ]\; = \;0, 
\end{eqnarray}
\begin{eqnarray}\label{4}									
&&s_b A_{\mu\nu\sigma} = \partial_\mu C_{\nu\sigma} + \partial_\nu C_{\sigma\mu}
+ \partial_\sigma C_{\mu\nu}, \qquad  s_b C_{\mu\nu} = \partial_\mu \beta_\nu
- \partial_\nu \beta_\mu, \qquad s_b \bar C_{\mu\nu} = B_{\mu\nu}, 
\nonumber\\
&& s_b A_{\mu} = \partial_\mu C, \quad s_b \bar C = B, \quad  
s_b \bar \beta_\mu = \bar F_\mu, \quad
s_b \beta_\mu = \partial_\mu C_2,  \quad s_b \bar B_{\mu\nu} = \partial_\mu f_\nu - \partial_\nu f_\mu, 
\nonumber\\
&& s_b {\bar C}_2 = B_5, \qquad s_b C_1 = - B_4, \qquad  s_b \bar C_1 = B_2, \qquad s_b \phi_\mu = f_\mu, \qquad s_b F_\mu = -\, \partial_\mu B_4,
\nonumber\\
&& s_b \bar f_\mu = \partial_\mu B_2, \; s_b  \Bigl [ C,\, C_2, \, B,\, B_1, \, B_2,\, B_3, \,  B_4,\, B_5,\, f_\mu,\, {\bar F}_\mu,\,\widetilde \phi_\mu,
B_{\mu\nu},  \, {\cal  B}_{\mu\nu}, \, \bar {\cal B}_{\mu\nu}
\Bigl ] = 0,
\end{eqnarray}
leave the action integral $S = \int d^4 x\,{\cal L}_{(B)}  $, corresponding to the total 
Lagrangian density ${\cal L}_{(B)} = {\cal L}_{(NG)} + {\cal L}_{(FP)} $,
invariant (i.e. $s_{(d)b} S = 0 $) because we observe the following:
\begin{eqnarray}\label{5}
s_d {\cal L}_{(B)} &=& -\,\dfrac{1}{2}\, \partial_\mu\, \Big[ (\partial^\mu\, \bar C^{\nu\sigma} + \partial^\nu\, \bar C^{\sigma\mu}
+ \partial^\sigma\, \bar C^{\mu\nu}) \,{\cal  B}_{\nu\sigma}  + {\cal B}^{\mu\nu}\, \bar F_\nu - B_4\, \partial^\mu\,\bar C_2  \nonumber\\
&+& B_3\, \bar F^\mu - B_5\, f^\mu - (\partial^\mu\, \bar \beta^\nu - \partial^\nu\, \bar \beta^\mu)\, f_\nu \Big] 
+ \partial_\mu \Big [B_1 \, \partial^\mu \bar C \Big ],
\end{eqnarray}
\begin{eqnarray}\label{6}
s_b\, {\cal L}_{(B)} &=& \dfrac{1}{2}\, \partial_\mu\, \Big[ (\partial^\mu\, C^{\nu\sigma} + \partial^\nu\, C^{\sigma\mu}
+ \partial^\sigma\, C^{\mu\nu}) \, B_{\nu\sigma}  + B^{\mu\nu}\, f_\nu - B_5\, \partial^\mu\, C_2  
\nonumber\\
&+&
B_2\, f^\mu + B_4\, \bar F^\mu - (\partial^\mu\,  \beta^\nu - \partial^\nu\, \beta^\mu)\,  \bar F_\nu \Big] 
+ \partial_\mu \Big [B \, \partial^\mu  C \Big ].
\end{eqnarray}
Corresponding to the Lagrangian density ${\cal L}_{(B)} = {\cal L}_{(NG)} + {\cal L}_{(FP)} $ [cf. Eqs. (1),(2)], we have a {\it coupled}
 Lagrangian density  ${\cal L}_{(\bar B)} = {\cal L}_{(ng)} + {\cal L}_{(fp)} $ whose non-ghost sector [i.e. ${\cal L}_{(ng)} $]
and FP-ghost sector [i.e. ${\cal L}_{(fp)} $]  (of the Lagrangian density ${\cal L}_{(\bar B)} $) are 																																						\begin{eqnarray}\label{7}
&&{\cal L}_{(ng)}  =  B \, (\partial \cdot A) - \dfrac{1}{2}\, B^2  + \dfrac{1}{2}\, B_2 \,\big (\partial \cdot \phi \big ) - \dfrac{1}{4}\, B_2^2  -
\dfrac{1}{2}\, B_3 \,\big (\partial \cdot \widetilde \phi \big ) + \dfrac{1}{4}\, B_3^2 
 \nonumber\\
&+&  \dfrac{1}{2} \, B_1^2 
+\, B_1\, \Big(\dfrac{1}{3!}\,\varepsilon^{\mu\nu\sigma\rho} \,\partial_\mu A_{\nu\sigma\rho} \Big) 
 - \,\dfrac{1}{4} \, \big( \bar B_{\mu\nu} \big)^2 - \dfrac{1}{2}\, \bar B_{\mu\nu} \, \Big[\partial_\sigma A^{\sigma\mu\nu}
+ \dfrac{1}{2}\, \big (\partial^\mu \phi^\nu - \partial^\nu \phi^\mu \big) \Big ]  
   \nonumber\\
&+& \dfrac{1}{4}\, \big( {\bar {\cal B}_{\mu\nu}} \big)^2 + \dfrac{1}{2} \, {\bar {\cal B}_{\mu\nu}}\, 
\Big[ \varepsilon^{\mu\nu\sigma\rho} \,\partial_\sigma A_\rho 
- \dfrac{1}{2}\, \big (\partial^\mu \widetilde \phi^\nu - \partial^\nu \widetilde \phi^\mu \big) \Big], 
 \end{eqnarray}																																					\begin{eqnarray}\label{8}
{\cal L}_{(fp)}  &=&  \dfrac{1}{2}\, \Big [\big (\partial_\mu \bar C_{\nu\sigma} +  \partial_\nu \bar C_{\sigma\mu} 
+ \partial_\sigma \bar C_{\mu\nu} \big ) \big (\partial^\mu C^{\nu\sigma} \big ) -
\big (\partial_\mu \bar C^{\mu\nu}  - \partial^\nu \bar C_1 \big ) F_\nu \nonumber\\ 
&+& \big (\partial_\mu  C^{\mu\nu}  - \partial^\nu  C_1 \big ) \bar f_\nu 
 +  \big( \partial \cdot \bar \beta \big) \, B_4
 - \big( \partial \cdot  \beta \big) \, B_5 - B_4 \, B_5 - 2\, \bar f^\mu\, F_\mu \nonumber\\
&-& \big (\partial_\mu \bar \beta_\nu - \partial_\mu \bar \beta_\nu \big ) 
\big (\partial^\mu \beta^\nu \big )  
-\, \partial_\mu \bar C_2\, \partial^\mu C_2  \Big ] + \partial_\mu \bar C\, \partial^\mu C, 
\end{eqnarray}
where a couple of {\it new} (i) bosonic auxiliary fields $(\bar B_{\mu\nu})\bar {\cal B}_{\mu\nu}$, and (ii) fermionic auxiliary fields $(F_\mu)\bar f_\mu$ have been invoked in the Lagrangian  densities ${\cal L}_{(ng)} $ and ${\cal L}_{(fp)}$,  respectively.	It is straightforward to check that the following off-shell nilpotent (i.e. $s_{ad}^2 = 0, \, s_{ab}^2 = 0 $) anti-dual-BRST ($s_{ad}$) and anti-BRST ($s_{ab} $) symmetry transformations																																						\begin{eqnarray}\label{9}
&& s_{ad} A_{\mu \nu\sigma} = \varepsilon_{\mu\nu\sigma\rho}\, \partial^\rho  C, 
\qquad
s_{ad} A_\mu = \dfrac{1}{2} \,\varepsilon_{\mu\nu\sigma\rho}\,\partial^\nu  C^{\sigma\rho}, \qquad
s_{ad}  C_{\mu\nu}  = -\, \big ( \partial_\mu \beta_\nu - \partial_\nu \beta_\mu \big ),  \nonumber\\
&& s_{ad} {\cal B}_{\mu\nu} = \partial_\mu F_\nu - \partial_\nu F_\mu, \qquad
s_{ad} \beta_\mu = -\, \partial_\mu C_2,  \qquad s_{ad} \bar \beta_\mu = +\, \bar f_\mu,  
\quad s_{ad} \widetilde \phi_\mu  = +\, F_\mu,   \nonumber\\
&&  s_{ad} \bar C_{\mu\nu} = + \, \bar {\cal B}_{\mu\nu},  \qquad s_{ad} \bar C = +\, B_1, \qquad
s_{ad} \bar C_2 = -\, B_5, \qquad s_{ad} \bar C_1 = -\, B_3,  \nonumber\\
&& s_{ad}  C_1 = -\,B_4, \qquad \quad s_{ad} f_\mu = -\, \partial_\mu B_4, \qquad s_{ad} \bar F_\mu = -\, \partial_\mu B_3, \nonumber\\
&&
s_{ad} \; \Bigl [ C,\, C_2,\,  B,\, B_1, \, B_2,\, B_3, \,  B_4,\, B_5,\, \bar f_\mu,\, {F}_\mu,\, \phi_\mu, \, B_{\mu\nu}, \, \bar B_{\mu\nu},\,
\bar {\cal B}_{\mu\nu} \Bigl ]\; = \;0, 
\end{eqnarray}
\begin{eqnarray}\label{10}
&&s_{ab} A_{\mu\nu\sigma} = \partial_\mu \bar C_{\nu\sigma} + \partial_\nu \bar C_{\sigma\mu}
+ \partial_\sigma \bar C_{\mu\nu}, \quad  s_{ab} \bar C_{\mu\nu} = \partial_\mu \bar \beta_\nu
- \partial_\nu \bar \beta_\mu, \quad s_{ab}  C_{\mu\nu} = \bar B_{\mu\nu},
\nonumber\\
&&
s_{ab} A_{\mu} = \partial_\mu \bar C, \quad s_{ab}  C = -\, B, \quad  
s_{ab}  \beta_\mu =  F_\mu, \quad
s_{ab} \bar \beta_\mu = \partial_\mu \bar C_2, \quad s_{ab} B_{\mu\nu} = \partial_\mu \bar f_\nu - \partial_\nu \bar f_\mu, 
\nonumber\\
&&
s_{ab} {C}_2 = B_4, \quad s_{ab} C_1 = - B_2, \quad  s_{ab} \bar C_1 = -\, B_5, 
\quad s_{ab} \phi_\mu = \bar f_\mu, \quad s_{ab} \bar F_\mu =-\, \partial_\mu B_5,
\nonumber\\
&& s_{ab} f_\mu = -\, \partial_\mu B_2, \;
 s_{ab}  \Bigl [ \bar C,\, \bar C_2,\, B,\, B_1,  B_2, B_3,   B_4, B_5,
  \bar f_\mu,\, {F}_\mu,
\,\widetilde \phi_\mu, \, \bar B_{\mu\nu}, \, 
 {\cal B}_{\mu\nu}, \, \bar {\cal B}_{\mu\nu} \Bigl ]  = 0, 
\end{eqnarray}
render the action integral $S = \int d^4 x\, {\cal L}_{\bar B}$ invariant (i.e. $s_{ad} S = 0, \, s_{ad} S = 0 $) because of the following 
transformations of ${\cal L}_{(\bar B)} $ under $s_{ad}$ and $s_{ab}$, namely; 																																					
\begin{eqnarray}\label{11}
s_{ad}\, {\cal L}_{(\bar B)} &=& \dfrac{1}{2}\, \partial_\mu\, \Big[ (\partial^\mu\,  C^{\nu\sigma} + \partial^\nu\,  C^{\sigma\mu}
+ \partial^\sigma\,  C^{\mu\nu}) \, \bar {\cal B}_{\nu\sigma}  - \bar {\cal B}^{\mu\nu}\, F_\nu  + B_5\, \partial^\mu\,  C_2  \nonumber\\
&-& B_3\, F^\mu + B_4\, \bar f^\mu - (\partial^\mu\,   \beta^\nu - \partial^\nu\,  \beta^\mu)\,  \bar f_\nu \Big] 
+ \partial_\mu \Big [B_1 \, \partial^\mu  C \Big ], 
\end{eqnarray}
\begin{eqnarray}\label{12}
s_{ab}\, {\cal L}_{(\bar B)} &=& \dfrac{1}{2}\, \partial_\mu\, \Big[\bar B^{\mu\nu}\, \bar f_\nu - \, (\partial^\mu\, \bar C^{\nu\sigma} + \partial^\nu\, \bar C^{\sigma\mu}
+ \partial^\sigma\, \bar C^{\mu\nu}) \, \bar B_{\nu\sigma}   + B_4\, \partial^\mu\, \bar C_2  
\nonumber\\
&-& B_5\, F^\mu + B_2\, \bar f^\mu - (\partial^\mu\,  \bar \beta^\nu - \partial^\nu\, \bar \beta^\mu)\,  F_\nu \Big] 
+ \partial_\mu \Big [B \, \partial^\mu  \bar C \Big ].
\end{eqnarray}
Thus, the action integral of our 4D BRST-quantized field-theoretic system respects total {\it four} off-shell nilpotent symmetry 
transformations [cf. Eqs. (3),(4),(9),(10)].																																																																		

We end our present 
section with the remarks that (i) the total kinetic terms for the Abelian 3-form and 1-form gauge fields
remain invariant under the nilpotent (anti-)BRST symmetry transformations, (ii) the total gauge-fixing terms, corresponding
to the Abelian 3-form and 1-form gauge fields, remain unchanged under the nilpotent
(anti-)dual-BRST transformations, and (iii) the BRST-quantized versions of the
Lagrangian densities [that exist for the free Abelian 3-form gauge theory in equations (1), (2), (7) and (8)] have been taken from our earlier
work where they have been systematically derived (see, e.g. [18]).


\section{Algebraic Structures: Discrete Duality and Continuous Nilpotent Symmetry Operators}

In our present section, we focus on (i) the the discrete duality symmetry transformation operators, (ii) the continuous and 
nilpotent symmetry transformation 
operators, and (iii) the algebraic structures that are satisfied by them. In particular, we attach a great of importance to the emergence of the
(anti-)BRST as well as the (anti-)co-BRST invariant CF-type restrictions in our present  discussion. In this context, first of all, we note that the 
following discrete duality symmetry transformations for the basic and auxiliary fields\footnote{It is worthwhile to mention that
our current discrete duality transformations (13) are slightly {\it different} from 
such transformations quoted in our earlier work [15] and {\it these}  are the correct transformations because
they satisfy {\it accurately} the algebraic structures that have been mentioned in equations (14) and (17).} 
\begin{eqnarray}\label{13}
&&A_\mu \longrightarrow \,\pm\, \dfrac{i}{3!}\, \varepsilon_{\mu\nu\sigma\rho}\, A^{\nu\sigma\rho}, \quad
A_{\mu\nu\sigma} \longrightarrow \,\pm\,i\, \varepsilon_{\mu\nu\sigma\rho}\, A^\rho, \quad B_{\mu\nu} \to \pm\, i\,{\cal B}_{\mu\nu}, \quad
{\cal B}_{\mu\nu} \to \pm\,  i\,\,B_{\mu\nu}, \nonumber\\
&& B \to \mp\, i\, B_1, \;\; B_1 \to \mp\, i\, B, \; \; B_2 \to \pm\,  i\, \,B_3, \;\; B_3 \to \pm\,i\,
 B_2, \;\; \phi_\mu \to \pm\,  i\, \widetilde \phi_\mu, \; \;
\widetilde \phi_\mu \to \pm\,  i\, \phi_\mu,
\nonumber\\
&& C_{\mu\nu} \longrightarrow \,\pm\,  i\, \bar C_{\mu\nu}, \quad \bar C_{\mu\nu} \longrightarrow \,\pm\,  i\, C_{\mu\nu}, \quad
\beta_\mu \to \pm\,  i\, \bar \beta_\mu, \quad \bar \beta_\mu \to \mp\, i\, \beta_\mu, \quad 
f_\mu \to \pm\,  i\, \bar F_\mu, \nonumber\\
&& \bar F_\mu \to \pm\,  i\,  f_\mu, \qquad B_4 \to \mp\,  i\, B_5, \qquad B_5 \to \pm\,  i\, B_4,
\qquad C \to  \pm\,  i\, \bar C, \qquad \bar C \to \pm\,  i\,  C, \nonumber\\
&&  C_2 \to \pm\,  i\,\bar C_2, \qquad 
\bar C_2 \to \pm\,  i\,  C_2, \qquad C_1 \to \pm\, i\, \bar C_1, \qquad 
\bar C_1 \to \pm\, i  \, C_1,
\end{eqnarray}
leave the Lagrangian density ${\cal L}_{(B)} = {\cal L}_{(NG)} + {\cal L}_{(FP)} $ [cf. Eqs. (1),(2)] invariant. It is worthwhile to mention, at this stage, 
that the interplay between the continuous symmetry operators [cf. Eq. (3),(4)] and the discrete symmetry operator [cf. Eq. (13)], provide the 
physical realization of the
well-known operator relationship: $\delta = \pm\, *\, d\, * $ between the (co-)exterior derivatives $(\delta)d$ of differential geometry because we 
observe the following
\begin{eqnarray}\label{14}
s_d \, \Sigma &=& +\, *\, s_b \, *\, \Sigma, \qquad \Sigma = \beta_\mu, \; \bar \beta_\mu, \, B_4,\,  B_5, 
\qquad  s_d \, \Phi = -\, *\, s_b \, *\, \Phi, \nonumber\\
\Phi &=& A_{\mu\nu\sigma}, \, B_{\mu\nu}, \, {\cal B}_{\mu\nu}, \, \bar C_{\mu\nu},\,  C_{\mu\nu}, \, A_\mu, 
\phi_\mu, \,  \widetilde \phi_\mu, \, f_\mu,\, \bar F_\mu, 
\nonumber\\
&& \bar C,  \, C, \,  \bar C_1, \,  C_1, \, \bar C_2, \,  C_2, \,  B, \, B_1,\,  B_2, \, B_3,  
\end{eqnarray}
where the $\pm$ signs in the relationships $s_d \, \Sigma = +\, *\, s_b \, *\, \Sigma $ and $s_d \, \Phi = -\, *\, s_b \, *\, \Phi $ are 
dictated by a couple of successive operations of the discrete duality symmetry transformations (13) 
on the generic fields $\Sigma $ and $\Phi $ which turn out to be the following (see, e.g. [19]):
\begin{eqnarray}\label{15}
 * \; \big (*\; \Sigma \big ) = +\, \Sigma, \;\;\qquad \;\;* \; \big (*\; \Phi \big ) = -\, \Phi.
 \end{eqnarray}
It is worthwhile to point out that the nilpotent (i.e. $\delta^2 = 0, \, d^2 = 0 $) (co-)exterior derivative operators  $(\delta)d$ are 
(i) physically realized  [cf. Eq. (14)]
in terms of the nilpotent (i.e. $s_d^2 = 0, \, s_b^2 = 0 $),  infinitesimal and continuous (dual-)BRST symmetry  transformation operators $s_{(d)b} $,
and (ii) the discrete duality symmetry transformations (13) provide the physical realizations of the Hodge duality $*$ operator
of differential geometry.

We now focus on the discrete duality symmetry transformations that exist for the Lagrangian density ${\cal L}_{(\bar B)} = {\cal L}_{(ng)} + {\cal L}_{(fp)} $
[cf. Eqs. (7),(8)]. We observe, in this connection, that {\it most} of the fields are common as far as the Lagrangian densities ${\cal L}_{(B)} $ 
and ${\cal L}_{(\bar B)} $ are concerned. These common basic and auxiliary fields
transform, under the discrete duality symmetry transformations for ${\cal L}_{(\bar B)} $, {\it same} as in (13). In the Lagrangian density  ${\cal L}_{(\bar B)} $,
there are {\it additional} Nakanishi-Lautrup type auxiliary fields (i.e. $\bar B_{\mu\nu}, \; \bar {\cal B}_{\mu\nu}, \;
\bar f_\mu, \; F_\mu $). These auxiliary fields transform,  under the discrete duality symmetry transofmfstions, as:
\begin{eqnarray}\label{16}
 \bar B_{\mu\nu} \to \pm\, i\, \bar {\cal B}_{\mu\nu}, \quad
\bar {\cal B}_{\mu\nu} \to \pm\, i\, \bar B_{\mu\nu},\quad F_\mu \to \pm\, i\, \bar f_\mu, \quad
\bar f_\mu \to \pm\, i\, F_\mu.
 \end{eqnarray}
Within the framework of BRST formalism, there is an alternative physical realization of the well-known mathematical relationship: $\delta = \pm\, *\, d\, * $ 
between the (co-)exterior derivatives $(\delta)d$ of differential geometry in terms of (i) the 
off-shell nilpotent (i.e. $s_{ab}^2  = 0, \; s_{ad}^2 = 0 $), infinitesimal and 
continuous anti-BRST symmetry transformations $s_{ab}$ [cf. Eq. (10)] and the
anti-co-BRST symmetry transformations $s_{ad}$ [cf. Eq. (9)], {\it and} (ii) the discrete duality\footnote{ We christen the discrete symmetry transformations
in (13) and (16) as the {\it duality} symmetry transformations because, under these discrete symmetry transformations, we note that (i) the gauge-fixing term
of the Abelian 1-form basic gauge field transforms to the kinetic term for the Abelian 3-form basic gauge field and vice-versa, and (ii) the gauge-fixing
term for the basic Abelian 3-form gauge field exchanges with the kinetic term for the Abelian 1-form basic gauge field. As stated earlier, the kinetic terms
and gauge-fixing terms for the basic Abelian 3-form and 1-form gauge fields owe their mathematical origins to the exterior 
and dual-exterior derivatives of differential geometry, respectively.}
symmetry transformations (13) and (16). This statement can be  
mathematically expressed in the following explicit form, namely; 
\begin{eqnarray}\label{17}
s_{ad} \, \Sigma &=& +\, *\, s_{ab} \, *\, \Sigma, \qquad \Sigma = \beta_\mu, \; \bar \beta_\mu, \, B_4,  \, B_5, \qquad  s_{ad} \, 
\widetilde \Phi = -\, *\, s_{ab} \, *\, \widetilde \Phi, \nonumber\\
\widetilde \Phi &=& A_{\mu\nu\sigma},   \,  \bar B_{\mu\nu},  \, \bar {\cal B}_{\mu\nu}, \, \bar C_{\mu\nu},  \, C_{\mu\nu}, \, A_\mu, 
\, \phi_\mu,  \widetilde \phi_\mu,  \, f_\mu, \, \bar F_\mu,
\nonumber\\
&& \bar C, \,  C,  \, \bar C_1,  \, C_1, \, \bar C_2,  \,  C_2,  \,  B,  \, B_1,  \, B_2, \,  B_3,  
\end{eqnarray}
where the $\pm$ signs in the relationships $s_{ad} \, \Sigma = +\, *\, s_{ab} \, *\, \Sigma $ and $s_{ad} \, 
\widetilde \Phi = +\, *\, s_{ab} \, *\, \widetilde \Phi $ are 
exactly the {\it same} as our observations in (15) with the replacement: $\Phi \to \widetilde \Phi $. The mathematical relationships in equations (14)
and (17) merely establish that the off-shell nilpotent (i.e. $s_{(a)b}^2 = 0,\; s_{(a)d}^2 = 0 $) (anti-)BRST $s_{(a)b} $ and (anti-)co-BRST $s_{(a)d} $ symmetry
transformation operators are {\it dual} to one-another (cf. Secs. 5 and 6 for more comments).

In our earlier works [16,17], we have been able to demonstrate that there are total {\it four} (anti-)BRST as well as the (anti-)co-BRST invariant
CF-type restrictions: $B_{\mu\nu} + \bar B_{\mu\nu} 
= (\partial_\mu  \phi_\nu  - \partial_\nu  \phi_\mu),\;
{\cal B}_{\mu\nu} + \bar {\cal B}_{\mu\nu} 
= (\partial_\mu \widetilde \phi_\nu  - \partial_\nu \widetilde \phi_\mu), \; f_\mu + F_\mu = \partial_\mu C_1, \;
\bar f_\mu + \bar F_\mu = \partial_\mu \bar C_1$ on our 4D BRST-quntized field-theoretic system. It is very interesting to point out that the following 
{\it non-trivial} anticommutators between the BRST and anti-BRST symmetry transformation operators exist (when they operate on the fields:
$A_{\mu\nu\sigma}, \, C_{\mu\nu}, \, \bar C_{\mu\nu} $), namely;
\begin{eqnarray}\label{18}
&& \{s_b, \,s_{ab}\}\, A_{\mu\nu\sigma} = \partial_\mu (B_{\nu\sigma} +  \bar B_{\nu\sigma}) + \partial_\nu (B_{\sigma\mu}
 +  \bar B_{\sigma\mu}) + \partial_\sigma (B_{\mu\nu} +  \bar B_{\mu\nu}), \nonumber\\
 && \{s_b, \,s_{ab}\}\, C_{\mu\nu} = \partial_\mu (f_\nu +  F_\nu) - \partial_\nu (f_\mu + F_\mu), \nonumber\\
  && \{s_b, \, s_{ab}\}\,\bar C_{\mu\nu} = \partial_\mu (\bar f_\nu +  \bar F_\nu) - \partial_\nu (\bar f_\mu + \bar F_\mu),
\end{eqnarray}
which turn out to be {\it zero} on the subspace of quantum fields where {\it three} of the above CF-type restrictions (i.e. 
$B_{\mu\nu} + \bar B_{\mu\nu} 
= (\partial_\mu  \phi_\nu  - \partial_\nu  \phi_\mu),\;
 \; f_\mu + F_\mu = \partial_\mu C_1, \;
\bar f_\mu + \bar F_\mu = \partial_\mu \bar C_1 $) 
are satisfied. It is pertinent to point out, at this juncture, that the absolute anticommutativity  (i.e. $\{ s_b, \; s_{ab} \} = 0 $) is
{\it trivially} satisfied for the {\it rest} of the fields of the coupled Lagrangian densities ${\cal L}_{(B)} $ 
and ${\cal L}_{(\bar B)} $. Against the backdrop of our present discussions, it is worthwhile to point out that the following
{\it non-trivial} anticommutators between the 
off-shell nilpotent co-BRST and anti-co-BRST symmetry transformation operators $s_{(a)d} $, namely;
\begin{eqnarray}\label{19}
&& \{s_d, \,s_{ad}\}\, A_{\mu} = +\, \dfrac{1}{2}\, \varepsilon_{\mu\nu\sigma\rho} \, \partial^\nu \big ({\cal B}^{\sigma\rho} 
+ \bar {\cal B}^{\sigma\rho} \big ), \nonumber\\
 && \{s_d, \,s_{ad}\}\, C_{\mu\nu} = + \, \big [\partial_\mu (f_\nu +  F_\nu) - \partial_\nu (f_\mu + F_\mu) \big ], \nonumber\\
  && \{s_b, \, s_{ad}\}\,\bar C_{\mu\nu} = + \, \big [\partial_\mu (\bar f_\nu +  \bar F_\nu) - \partial_\nu (\bar f_\mu + \bar F_\mu) \big ],
\end{eqnarray}
turn out to be zero provided we use the {\it three} specific CF-type restrictions
(i.e. ${\cal B}_{\mu\nu} + \bar {\cal B}_{\mu\nu} 
= (\partial_\mu \widetilde \phi_\nu  - \partial_\nu \widetilde \phi_\mu), \; f_\mu + F_\mu = \partial_\mu C_1, \;
\bar f_\mu + \bar F_\mu = \partial_\mu \bar C_1 $) which define the subspace of qauntum fields where the 
co-BRST and anti-co-BRST symmetry transformation operators absolutely anticommute (i.e. $\{ s_d, \; s_{ad} \} = 0 $).
The absolute anticommutativity (i.e. $\{ s_d, \; s_{ad} \} = 0 $)
is satisfied {\it trivially} for the rest the quantum fields of our 4D BRST-quantized theory. Hence, we finally conclude that the 
absolute anticommutativity (i.e. $\{ s_b, \; s_{ab} \} = 0 $ and $\{ s_d, \; s_{ad} \} = 0$) between the off-shell nilpotent
(i.e. $s_{(a)b}^2 = 0, \, s_{(a)d}^2 = 0 $)
(anti-)BRST ($s_{(a)b} $)  and (anti-)dual-BRST ($s_{(a)d} $) symmetry transformation operators 
 are satisfied
provided we utilize the {\it specific}  set of only\footnote{It has been shown in our earlier works [16,17] that the requirements of the {\it simultaneous}
(anti-)BRST and (anti-)co-BRST invariance(s) of the coupled Lagrangian densities ${\cal L}_{(B)} $ and ${\cal L}_{(\bar B)} $ {\it also} lead
to the derivations of {\it only} three specific sets of the CF-type restrictions (out of the total {\it four}).} 
{\it three} CF-type restrictions (out of the total {\it four}).

We end this section with a few key concluding remarks. First of all, we observe that the following algebraic operator structures
(in the terminology of the anticommutators)
\begin{eqnarray}\label{20}
 \big \{ s_b, \; s_{ad} \big \} = 0, \qquad \big \{ s_d, \; s_{ab} \big \} = 0,
\end{eqnarray}
are satisfied modulo a set of U(1) {\it vector} gauge symmetry-type transformations
(cf. Sec. 4 below for more discussions on this issue). Second, the following non-trivial anticommutators between the {\it appropriate}  nilpotent 
(anti-)BRST and (anti-)co-BRST symmetry operators
\begin{eqnarray}\label{21}
 \big \{ s_b, \; s_{d} \big \} = s_\omega, \;\;\qquad \;\;\big \{ s_{ad}, \; s_{ab} \big \} = s_{\bar \omega},
\end{eqnarray}
define the {\it two} non-trivial bosonic symmetry transformation operators $s_\omega $ and $s_{\bar \omega} $, respectively, for our 4D BRST-quantized field-theoretic system. Third, we would like to point out that some of the key relationships between the co-BRST and BRST 
symmetry transformations have {\it not} been captured in
equation (14). Taking into account the total discrete duality symmetry transformations (13) and (16), it is straightforward to check that: 
\begin{eqnarray}\label{22}
&&s_d \bar f_\mu = -\, *\, s_b\, * \bar f_\mu = \partial_\mu B_5, \qquad s_d F_\mu = -\, *\, s_b\, * \, F_\mu = +\, \partial_\mu B_3,
\nonumber\\
&& s_d \bar {\cal B}_{\mu\nu} = -\, *\, s_b\, * \,\bar {\cal B}_{\mu\nu} = (\partial_\mu \bar F_\nu - \partial_\nu \bar F_\mu).
\end{eqnarray}
Finally, it is worthwhile to mention that, even though the mathematical relationships in equations (14) and (17) are {\it similar}, it is the ghost number 
considerations\footnote{It is straightforward to note that 
the pair of nilpotent operators ($s_d, \;  s_{ab}$) lower the ghost number of a field by {\it one} on which they operate. On the contrary, the pair of
nilpotent operators ($s_b, \;  s_{ad}$) raise the
ghost number of a field by {\it one} (on which theory operate). These observations are similar to the consequences that emerge out when 
the nilpotent (co-)exterior derivatives operate on a given differential form.} 
that decide, out of the pairs ($s_d, \; s_{ad} $) and ($s_b, \; s_{ab} $), which operator corresponds to the nilpotent (i.e. $\delta^2 = 0 $) 
co-exterior derivative $\delta $ and which
operator represents the nilpotent (i.e. $d^2 = 0 $) 
exterior derivative $d$ of differential geometry (cf. Secs. 5 and 6 for more discussions).


\section{Bosonic Symmetry Transformations: Uniqueness Property and Full Set of CF-Type Restrictions}

The non-nilpotent (i.e. $s_\omega^2 \neq 0 $) bosonic 
symmetry operator $s_\omega$ [cf. Eq. (20)] can operate on the  
{\it individual} fields of this Lagrangian density  ${\cal L}_{(B)} $. The following infinitesimal, continuous and non-nilpotent 
symmetry transformations
(under $s_\omega = \{ s_b, \; s_d \}$), namely;
\begin{eqnarray}\label{23}
&& s_\omega A_{\mu\nu\sigma} =  \varepsilon_{\mu\nu\sigma\rho}\, \partial^\rho B  +
\Big (\partial_\mu\, {\cal B}_{\nu\sigma} + \partial_\nu\, {\cal B}_{\sigma\mu}
+ \partial_\sigma\, {\cal B}_{\mu\nu} \Big ),   \nonumber\\
&& s_\omega A_\mu = \dfrac{1}{2} \, \varepsilon_{\mu\nu\sigma\rho}\, \partial^\nu B^{\sigma\rho} - \partial_\mu B_1, 
\qquad s_\omega \beta_\mu = +\,\partial_\mu  B_4,
\qquad s_\omega \bar \beta_\mu = +\,\partial_\mu B_5,  \nonumber\\
&& s_\omega C_{\mu\nu} = -\, \big (\partial_\mu f_\nu - \partial_\nu f_\mu \big ), \qquad 
s_\omega \bar C_{\mu\nu} = +\, \big (\partial_\mu \bar F_\nu - \partial_\nu \bar F_\mu \big ), \nonumber\\
&& s_\omega \Big[B, \, B_1,  B_2,  B_3,  B_4,  B_5, \, \phi_\mu,\, \widetilde \phi_\mu,\, f_\mu, \, \bar F_\mu,\, 
C, \bar C, \, C_1,  \bar C_1,  C_2,  \bar C_2, B_{\mu\nu}, \, {\cal B}_{\mu\nu}  \Big] = 0,
 \end{eqnarray}
lead to the transformation of the Lagrangian density ${\cal L}_{(B)} $ as 
\begin{eqnarray}\label{24}
&&s_\omega \, {\cal L}_{(B)} = 
 \dfrac{1}{2}\, \partial_\mu\, \Big[ \big \{ \partial^\mu\,{\cal B}^{\nu\sigma} + \partial^\nu\,{\cal B}^{\sigma\mu}
+ \partial^\sigma\, {\cal B}^{\mu\nu} \big \} \, B_{\nu\sigma} \nonumber\\
&&- \big \{\partial^\mu\, B^{\nu\sigma} + \partial^\nu\, B^{\sigma\mu}
+ \partial^\sigma\, B^{\mu\nu} \big \} \, {\cal B}_{\nu\sigma}  
+ B_4 \, \partial^\mu B_5 - B_5 \, \partial^\mu B_4 \nonumber\\
&& + \,\big (\partial^\mu f^\nu - \partial^\nu f^\mu \big )\, \bar F_\nu + \big (\partial^\mu \bar F^\nu - \partial^\nu \bar F^\mu \big )\, f_\nu
\Big ] + \partial_\mu \Big [\big (B_1\, \partial^\mu B - B\, \partial^\mu B_1 \big ) \Big ],
\end{eqnarray}
thereby rendering the action integral $S = \int d^4x\, {\cal L}_{(B)} $ invariant (i.e. $s_\omega S = 0 $). The transformation in (24) can be easily derived
by using our observations in equations (5) and (6). In other words, we have $ s_\omega {\cal L}_{(B)} = \big ( s_b \, s_d + s_d \, s_b \big )\, {\cal L}_{(B)} $ 
which implies the transformation (24).

Against the backdrop of the above discussions, we note that the operation of the bosonic symmetry transformation operator
$s_{\bar\omega} = \{ s_{ab}, \; s_{ad} \}$ on the individual fields of the Lagrangian density ${\cal L}_{(\bar B)} $ leads to the following:
\begin{eqnarray}\label{25}
&& s_{\bar \omega} A_{\mu\nu\sigma} =  -\,\varepsilon_{\mu\nu\sigma\rho}\, \partial^\rho B  +
\Big (\partial_\mu\, \bar {\cal B}_{\nu\sigma} + \partial_\nu\, \bar {\cal B}_{\sigma\mu}
+ \partial_\sigma\, \bar {\cal B}_{\mu\nu} \Big ),   \nonumber\\
&& s_{\bar \omega} A_\mu = \dfrac{1}{2} \, \varepsilon_{\mu\nu\sigma\rho}\, \partial^\nu \bar B^{\sigma\rho} + \partial_\mu B_1, 
\qquad s_{\bar \omega} \beta_\mu = -\, \partial_\mu  B_4,
\qquad s_{\bar \omega} \bar \beta_\mu = -\, \partial_\mu B_5,  \nonumber\\
&& s_{\bar \omega} C_{\mu\nu} = -\, \big (\partial_\mu F_\nu - \partial_\nu F_\mu \big ), \qquad 
s_{\bar \omega} \bar C_{\mu\nu} = +\, \big (\partial_\mu \bar f_\nu - \partial_\nu \bar f_\mu \big ), \nonumber\\
&& s_{\bar \omega} \Big[B, \, B_1,  B_2,  B_3,  B_4,  B_5, \, \phi_\mu,\, \widetilde \phi_\mu,\, \bar f_\mu, \, F_\mu,\, 
C, \bar C, \, C_1,  \bar C_1,  C_2,  \bar C_2, \bar B_{\mu\nu}, \, \bar {\cal B}_{\mu\nu}  \Big] = 0.
 \end{eqnarray}
Taking into account the sum of transformations (23) and (25), we obtain:
\begin{eqnarray}\label{26}
&& \big (s_\omega + s_{\bar \omega} \big ) A_{\mu\nu\sigma} =    +\, \Big [
\partial_\mu\, \big ({\cal B}_{\nu\sigma} + \bar {\cal B}_{\nu\sigma} \big )+ \partial_\nu\, \big ({\cal B}_{\sigma\mu} + \bar {\cal B}_{\sigma\mu} \big )
+ \partial_\sigma\, \big ({\cal B}_{\mu\nu} + \bar {\cal B}_{\mu\nu} \big ) \Big ],   \nonumber\\
&&\big (s_\omega + s_{\bar \omega} \big ) A_\mu = \dfrac{1}{2} \, \varepsilon_{\mu\nu\sigma\rho}\, \partial^\nu \big (B^{\sigma\rho} + \bar B^{\sigma\rho} \big ), 
\nonumber\\
&& \big (s_\omega + s_{\bar \omega} \big ) C_{\mu\nu} = -\, \Big [\big (\partial_\mu (f_\nu + F_\nu ) - \partial_\nu (f_\mu + F_\mu )\big ) \Big ], \nonumber\\ 
&& \big (s_\omega + s_{\bar \omega} \big )  \bar C_{\mu\nu} = +\, \Big [ \big (\partial_\mu (\bar f_\nu + \bar F_\nu)  
- \partial_\nu (\bar f_\mu + \bar F_\mu) \big ) \Big ], \nonumber\\
&& \big (s_\omega + s_{\bar \omega} \big ) \Big[ B, \, B_1,  B_2,  B_3,  B_4,  B_5,  \phi_\mu, \widetilde \phi_\mu, f_\mu,  \bar F_\mu,  
\bar f_\mu, F_\mu, \beta_\mu, \bar \beta_\mu,\nonumber\\
&& C, \bar C, \, C_1,  \bar C_1,  C_2,  \bar C_2, B_{\mu\nu}, \, {\cal B}_{\mu\nu},   \bar B_{\mu\nu}, \, \bar {\cal B}_{\mu\nu}  \Big] = 0.
 \end{eqnarray}
It is crystal clear that the operator relationship $s_\omega \,+ \, s_{\bar\omega} = 0$ is {\it true} provided we consider our whole BRST-quantized theory on the
subspace of quantum fields where {\it all} the {\it four} CF-type restrictions (i.e. $B_{\mu\nu} + \bar B_{\mu\nu} 
= (\partial_\mu  \phi_\nu  - \partial_\nu  \phi_\mu),\; {\cal B}_{\mu\nu} + \bar {\cal B}_{\mu\nu} 
= (\partial_\mu \widetilde \phi_\nu  - \partial_\nu \widetilde \phi_\mu), \; f_\mu + F_\mu = \partial_\mu C_1, \;
\bar f_\mu + \bar F_\mu = \partial_\mu \bar C_1$) are satisfied {\it together}. This observation should be contrasted against the 
backdrop of our observations in the
context of the proof of the absolute anticommutativity (i) between the nilpotent BRST and anti-BRST symmetry transformation operators
[cf. Eq. (18)], and (ii) between the nilpotent co-BRST and anti-co-BRST symmetry transformation operators [cf. Eq. (19)], where
{\it only} three (out of the total {\it four})
specific types of CF-type restrictions  have been invoked. We conclude, ultimately, that there is a {\it unique} bosonic symmetry transformation
operator (e.g. $s_\omega $) in our theory.

As far as the invariance of the action integral $S = \int d^4 x\, {\cal L}_{(\bar B)} $ (corresponding to the Lagrangian density 
${\cal L}_{(\bar B)}$) w.r.t the bosonic symmetry transformation operator $s_{\bar \omega} $ is concerned, we observe that 
${\cal L}_{(\bar B)}$ transforms to the following total spacetime derivative, namely;
\begin{eqnarray}\label{27}
&&s_{\bar \omega} \, {\cal L}_{(\bar B)} = 
\dfrac{1}{2}\, \partial_\mu\, \Big[ \big \{\partial^\mu\, \bar B^{\nu\sigma} + \partial^\nu\, \bar B^{\sigma\mu}
+ \partial^\sigma\, \bar B^{\mu\nu} \big \} \, \bar {\cal B}_{\nu\sigma}  \nonumber\\
&& - \,\big \{ \partial^\mu\, \bar {\cal B}^{\nu\sigma} + \partial^\nu\, \bar {\cal B}^{\sigma\mu}
+ \partial^\sigma\, \bar {\cal B}^{\mu\nu} \big \} \, \bar B_{\nu\sigma} + B_5 \, \partial^\mu B_4  - B_4 \, \partial^\mu B_5 \nonumber\\
&& - \,\big (\partial^\mu \bar f^\nu - \partial^\nu \bar f^\mu \big )\,  F_\nu - \big (\partial^\mu  F^\nu - \partial^\nu  F^\mu \big )\, \bar f_\nu
\Big ] + \partial_\mu \Big [\big (B\, \partial^\mu B_1 - B_1\, \partial^\mu B \big ) \Big ].
\end{eqnarray}
Hence, using the Gauss divergence theorem, it is clear that the action integral respects (i.e. $s_{\bar \omega} S = 0 $)
the bosonic symmetry transformations ($s_{\bar\omega} $) that have been listed in equation (25).
In addition to our observations in equations (24) and (27), we note that the following bosonic 
symmetry transformation of the Lagrangian density ${\cal L}_{(\bar B)} $  is {\it also} true, namely;
\begin{eqnarray}\label{28}
&&s_\omega \, {\cal L}_{(\bar B)} = +\,\partial_\mu \Big [\big (B_1\, \partial^\mu B  - B\, \partial^\mu B_1) \Big ]
+  \dfrac{1}{2}\, \partial_\mu\, \Big[ \big \{ \partial^\mu\,B^{\nu\sigma} + \partial^\nu\,B^{\sigma\mu}
+ \partial^\sigma\, B^{\mu\nu} \big \} \,\bar {\cal  B}_{\nu\sigma} \nonumber\\
&&- \big \{\partial^\mu\, {\cal B}^{\nu\sigma} + \partial^\nu\,{\cal B}^{\sigma\mu}
+ \partial^\sigma\, {\cal B}^{\mu\nu} \big \} \, \bar B_{\nu\sigma}  
+ B_4 \, \partial^\mu B_5 - B_5 \, \partial^\mu B_4 
- \,\big (\partial^\mu  f^\nu - \partial^\nu  f^\mu \big )\, \bar f_\nu \nonumber\\
&& - \big (\partial^\mu \bar F^\nu - \partial^\nu \bar F^\mu \big )\, F_\nu \Big ] 
+ \dfrac{1}{2} \Big [\big (\partial^\mu \bar F^\nu - \partial^\nu \bar F^\mu \big )\,\partial_\mu F_\nu 
+ \big (\partial^\mu f^\nu - \partial^\nu f^\mu \big )\,\partial_\mu \bar f_\nu \nonumber\\
&& + \big \{\partial^\mu\, {\cal B}^{\nu\sigma} + \partial^\nu\,{\cal B}^{\sigma\mu}
+ \partial^\sigma\, {\cal B}^{\mu\nu} \big \} \, \partial_\mu \bar B_{\nu\sigma}
- \big \{\partial^\mu\, B^{\nu\sigma} + \partial^\nu\, B^{\sigma\mu}
+ \partial^\sigma\, B^{\mu\nu} \big \} \, \partial_\mu \bar {\cal B}_{\nu\sigma} \Big ],
\end{eqnarray}
where the infinitesimal and  continuous bosonic symmetry transformations ($s_\omega $) are listed in equation (23). On the other hand, 
the application of the bosonic symmetry transformation operator $s_{\bar\omega} $ [cf. Eq. (25)] on the Lagrangian density ${\cal L}_{(B)} $ leads to the following:
\begin{eqnarray}\label{29}
&&s_{\bar \omega} \, {\cal L}_{(B)} = +\, \partial_\mu \Big [\big (B\, \partial^\mu B_1 - B_1\, \partial^\mu B \big ) \Big ] 
+ \dfrac{1}{2}\, \partial_\mu\, \Big[ \big \{\partial^\mu\, \bar {\cal B}^{\nu\sigma} + \partial^\nu\, \bar{\cal B}^{\sigma\mu}
+ \partial^\sigma\, \bar{\cal B}^{\mu\nu} \big \} \, B_{\nu\sigma}  \nonumber\\
&& - \,\big \{ \partial^\mu\, \bar B^{\nu\sigma} + \partial^\nu\, \bar B^{\sigma\mu}
+ \partial^\sigma\, \bar B^{\mu\nu} \big \} \, {\cal B}_{\nu\sigma} + B_5 \, \partial^\mu B_4 - B_4 \, \partial^\mu B_5  
 + \,\big (\partial^\mu \bar f^\nu - \partial^\nu \bar f^\mu \big )\,  f_\nu \nonumber\\
&&+ \big (\partial^\mu  F^\nu - \partial^\nu  F^\mu \big )\, \bar F_\nu
\Big ] - \dfrac{1}{2} \Big [\big (\partial^\mu F^\nu - \partial^\nu F^\mu \big )\,\partial_\mu \bar F_\nu 
+ \big (\partial^\mu \bar f^\nu - \partial^\nu \bar f^\mu \big )\,\partial_\mu f_\nu \nonumber\\
&& + \big \{\partial^\mu\, \bar {\cal B}^{\nu\sigma} + \partial^\nu\, \bar {\cal B}^{\sigma\mu}
+ \partial^\sigma\, \bar {\cal B}^{\mu\nu} \big \} \, \partial_\mu  B_{\nu\sigma}
- \big \{\partial^\mu\, \bar B^{\nu\sigma} + \partial^\nu\, \bar B^{\sigma\mu}
+ \partial^\sigma\, \bar B^{\mu\nu} \big \} \, \partial_\mu  {\cal B}_{\nu\sigma} \Big ].
\end{eqnarray}
All the terms, that are outside the total spacetime derivatives, in the above equations (28) and (29), can be expressed in terms of 
{\it all} the four CF-type restrictions on our theory. To be precise, using a slightly involved algebraic exercise, we can recast
the above equation (28), in terms of the CF-type restrictions, as follows:
\begin{eqnarray}\label{30}
&&s_\omega \, {\cal L}_{(\bar B)} = +\,\partial_\mu \Big [\big (B_1\, \partial^\mu B  - B\, \partial^\mu B_1) \Big ]
+  \dfrac{1}{2}\, \partial_\mu\, \Big[ \big \{ \partial^\mu\,B^{\nu\sigma} + \partial^\nu\,B^{\sigma\mu}
+ \partial^\sigma\, B^{\mu\nu} \big \} \,\bar {\cal  B}_{\nu\sigma} \nonumber\\
&&- \big \{\partial^\mu\, {\cal B}^{\nu\sigma} + \partial^\nu\,{\cal B}^{\sigma\mu}
+ \partial^\sigma\, {\cal B}^{\mu\nu} \big \} \, \bar B_{\nu\sigma}  
+ B_4 \, \partial^\mu B_5 - B_5 \, \partial^\mu B_4 
- \,\big (\partial^\mu  f^\nu - \partial^\nu  f^\mu \big )\, \bar f_\nu \nonumber\\
&& - \big (\partial^\mu \bar F^\nu - \partial^\nu \bar F^\mu \big )\, F_\nu \Big ] 
+ \frac{1}{2}\, \big [\partial^\mu  \{ f^\nu + F^\nu - \partial^\nu C_1 \}
 - \partial^\nu  \{ f^\mu + F^\mu - \partial^\mu C_1 \} \big ] \,\partial_\mu \bar f_\nu \nonumber\\
&& + \frac{1}{2}\, \big [\partial^\mu  \{ \bar f^\nu + \bar F^\nu - \partial^\nu \bar C_1 \}
 - \partial^\nu  \{ \bar f^\mu + \bar F^\mu - \partial^\mu \bar C_1 \} \big ] \,\partial_\mu F_\nu 
- \frac{1}{2}\, \big [  \partial^\mu \{ B^{\nu\sigma} + \bar B^{\nu\sigma} \nonumber\\
&&- (\partial^\nu \phi^\sigma - \partial^\sigma \phi^\nu )  \} 
+ \,\partial^\nu  \{ B^{\sigma\mu} + \bar B^{\sigma\mu} 
- (\partial^\sigma \phi^\mu - \partial^\mu \phi^\sigma )  \} 
+ \partial^\sigma  \{ B^{\mu\nu} + \bar B^{\mu\nu} \nonumber\\
&&- (\partial^\mu \phi^\nu - \partial^\nu \phi^\mu )  \} \big ]\, \partial_\mu \bar {\cal B}_{\nu\sigma} 
+ \frac{1}{2}\, \big [  \partial^\mu \{ {\cal B}^{\nu\sigma} + \bar{\cal B}^{\nu\sigma} 
- (\partial^\nu \widetilde \phi^\sigma - \partial^\sigma \widetilde \phi^\nu )  \} 
+ \,\partial^\nu  \{{\cal B}^{\sigma\mu} + \bar {\cal B}^{\sigma\mu} \nonumber\\
&&- (\partial^\sigma \widetilde \phi^\mu - \partial^\mu \widetilde \phi^\sigma )  \} + \partial^\sigma  \{ {\cal B}^{\mu\nu} + \bar {\cal B}^{\mu\nu} 
- (\partial^\mu \widetilde \phi^\nu - \partial^\nu \widetilde \phi^\mu )  \} \big ]\, \partial_\mu \bar B_{\nu\sigma}.
\end{eqnarray}
A close and careful look at the above equation demonstrates that, if we invoke the validity of all the {\it four} CF-type restrictions on our theory,
we find that the r.h.s. of the above equation becomes a total spacetime derivative. In other words, the action integral (corresponding to the
Lagrangian density ${\cal L}_{(\bar B)} $) respects [cf. Eqs. (27),(30)] both the 
bosonic transformations  $s_{\bar \omega} $ and $s_\omega $ on the subspace of quantum fields where all the {\it four} CF-type restrictions are satisfied.
Against the backdrop of the above discussions, we observe that
equation (29) can be recast, following exactly similar kind of algebraic exercise, in the following form:
\begin{eqnarray}\label{31}
&&s_{\bar \omega} \, {\cal L}_{(B)} = +\, \partial_\mu \Big [\big (B\, \partial^\mu B_1 - B_1\, \partial^\mu B \big ) \Big ] 
+ \dfrac{1}{2}\, \partial_\mu\, \Big[ \big \{\partial^\mu\, \bar {\cal B}^{\nu\sigma} + \partial^\nu\, \bar{\cal B}^{\sigma\mu}
+ \partial^\sigma\, \bar{\cal B}^{\mu\nu} \big \} \, B_{\nu\sigma}  \nonumber\\
&& - \,\big \{ \partial^\mu\, \bar B^{\nu\sigma} + \partial^\nu\, \bar B^{\sigma\mu}
+ \partial^\sigma\, \bar B^{\mu\nu} \big \} \, {\cal B}_{\nu\sigma} + B_5 \, \partial^\mu B_4 - B_4 \, \partial^\mu B_5  
 + \,\big (\partial^\mu \bar f^\nu - \partial^\nu \bar f^\mu \big )\,  f_\nu \nonumber\\
&&+ \big (\partial^\mu  F^\nu - \partial^\nu  F^\mu \big )\, \bar F_\nu
\Big ] - \frac{1}{2}\;
\big (\partial^\mu\, B^{\nu\sigma} + \partial^\nu\,B^{\sigma\mu}
+ \partial^\sigma\, B^{\mu\nu} \big )\, \partial_\mu \big [{\cal B}_{\nu\sigma} + \bar {\cal B}_{\nu\sigma} - (\partial_\mu \widetilde \phi_\nu 
- \partial_\nu \widetilde \phi_\mu)\big ]
\nonumber\\
&& + \frac{1}{2}\, \big (\partial^\mu \bar F^\nu - \partial^\nu \bar F^\mu \big ) \, \partial_\mu \big [ f_\nu +  F_\nu - \partial_\nu  C_1 \big ]
+ \frac{1}{2}\, \big [  \partial^\mu \{ B^{\nu\sigma} + \bar B^{\nu\sigma} - (\partial^\nu \phi^\sigma - \partial^\sigma \phi^\nu )  \} \nonumber\\
&&+ \,\partial^\nu  \{ B^{\sigma\mu} + \bar B^{\sigma\mu} - (\partial^\sigma \phi^\mu - \partial^\mu \phi^\sigma )  \} 
+ \partial^\sigma  \{ B^{\mu\nu} + \bar B^{\mu\nu} - (\partial^\mu \phi^\nu - \partial^\nu \phi^\mu )  \} \big ]\, \partial_\mu {\cal B}_{\nu\sigma}
\nonumber\\
&& - \frac{1}{2}\, \big [\partial^\mu  \{\bar f^\nu + \bar F^\nu - \partial^\nu \bar C_1  \} 
- \partial^\nu \big \{\bar f^\mu + \bar F^\mu- \partial^\mu \bar C_1 \big \} \big ]\, \partial_\mu f_\nu.
\end{eqnarray}
We note, once again, that if we exploit the validity of  the {\it all} the four CF-type restrictions on our theory, we observe that the 
r.h.s. of equation (31) becomes a total spacetime derivative. In other words, the 
action integral, corresponding to the Lagrangian density ${\cal L}_{(B)} $,  respects [cf. Eqs. (24),(31)]
both the bosonic symmetry transformations  $s_{\omega} $ and $s_{\bar \omega} $ on the subspace of quantum 
fields where all the {\it four} CF-type restrictions are satisfied.
Thus, in the language of the symmetry considerations, we have   
established that the coupled (but equivalent) Lagrangian densities ${\cal L}_{(B)} $ 
and ${\cal L}_{(\bar B)} $ respect {\it both} the bosonic symmetry transformations $s_\omega $ and $s_{\bar \omega}$
 provided {\it all} the four CF-type restrictions are satisfied {\it together}.

We would like to add here that the
operator equation: $(s_\omega + s_{\bar \omega}) = 0 $ [cf. Eq. (26)], is satisfied at the level of 
the transformations of the {\it coupled} Lagrangian densities ${\cal L}_{(B)} $  
and ${\cal L}_{(\bar B)} $, too, under the bosonic symmetry transformations [cf. Eqs. (23),(25)].  In this context, 
taking into account our observations in equations (27) and  (30), 
it is very interesting to point out that the following is {\it also} true:
\begin{eqnarray}\label{32}
&&\big (s_\omega + s_{\bar \omega} \big ) \, {\cal L}_{(\bar B)} = 
 \dfrac{1}{2}\, \partial_\mu\, \Big[ \Big (\partial^\mu \{ B^{\nu\sigma} + \bar B^{\nu\sigma} - (\partial^\nu \phi^\sigma 
- \partial^\sigma \phi^\nu) \} + \partial^\nu \{ B^{\sigma\mu} + \bar B^{\sigma\mu} \nonumber\\
&& - (\partial^\sigma \phi^\mu
- \partial^\mu \phi^\sigma) \}  
+\partial^\sigma \{ B^{\mu\nu} + \bar B^{\mu\nu} - (\partial^\mu \phi^\nu 
- \partial^\nu \phi^\mu) \} \Big ) \, \bar {\cal B}_{\nu\sigma} 
- \Big (\partial^\mu \{ {\cal B}^{\nu\sigma} + \bar{\cal B}^{\nu\sigma} \nonumber\\
&&- (\partial^\nu \widetilde \phi^\sigma 
- \partial^\sigma \widetilde \phi^\nu) \} 
+ \partial^\nu \{{\cal B}^{\sigma\mu} + \bar {\cal B}^{\sigma\mu} - (\partial^\sigma \widetilde \phi^\mu
- \partial^\mu \widetilde \phi^\sigma) \} 
+\partial^\sigma \{ {\cal B}^{\mu\nu} + \bar {\cal B}^{\mu\nu} \nonumber\\
&&- (\partial^\mu \widetilde \phi^\nu 
- \partial^\nu \widetilde \phi^\mu) \} \Big )  \bar B_{\nu\sigma} 
-\,\big (\partial^\mu  \{\bar f^\nu + \bar F^\nu - \partial^\nu \bar C_1  \} 
- \partial^\nu \big \{\bar f^\mu + \bar F^\mu  
- \partial^\mu \bar C_1 \big \} \big )\, F_\nu \nonumber\\
&&- \,\big (\partial^\mu  \{ f^\nu + F^\nu 
- \partial^\nu C_1 \} 
 - \partial^\nu  \{ f^\mu + F^\mu - \partial^\mu C_1 \} \big ) \, \bar f_\nu  \Big ] 
+ \dfrac{1}{2}\, \big (\partial^\mu  \{ f^\nu +  F^\nu - \partial^\nu  C_1  \} \nonumber\\
&&- \partial^\nu \big \{ f^\mu +  F^\mu- \partial^\mu  C_1 \big \} \big )\, \partial_\mu  \bar f_\nu 
 + \dfrac{1}{2}\, \big (\partial^\mu  \{\bar f^\nu + \bar F^\nu - \partial^\nu \bar C_1  \} 
- \partial^\nu \big \{\bar f^\mu + \bar F^\mu- \partial^\mu \bar C_1 \big \} \big )\, \partial_\mu   F_\nu \nonumber\\
 &&- \dfrac{1}{2}\, \Big [\Big (\partial^\mu \{ B^{\nu\sigma} + \bar B^{\nu\sigma} 
- (\partial^\nu \phi^\sigma 
- \partial^\sigma \phi^\nu) \} 
+ \partial^\nu \{ B^{\sigma\mu} + \bar B^{\sigma\mu} - (\partial^\sigma \phi^\mu
- \partial^\mu \phi^\sigma) \} \nonumber\\
&&+\partial^\sigma \{ B^{\mu\nu} + \bar B^{\mu\nu} 
- (\partial^\mu \phi^\nu 
- \partial^\nu \phi^\mu) \} \Big ) \Big ]\, \partial_\mu  \bar {\cal B}_{\nu\sigma} 
+ \dfrac{1}{2}\, \Big [\Big (\partial^\mu \{ {\cal B}^{\nu\sigma} + \bar{\cal B}^{\nu\sigma} 
- (\partial^\nu \widetilde \phi^\sigma 
- \partial^\sigma \widetilde \phi^\nu) \} \nonumber\\
&&+ \partial^\nu \{{\cal B}^{\sigma\mu} + \bar {\cal B}^{\sigma\mu} - (\partial^\sigma \widetilde \phi^\mu
- \partial^\mu \widetilde \phi^\sigma) \} 
+\partial^\sigma \{ {\cal B}^{\mu\nu} + \bar {\cal B}^{\mu\nu} - (\partial^\mu \widetilde \phi^\nu 
- \partial^\nu \widetilde \phi^\mu) \} \Big ) \Big ]\, \partial_\mu \bar B_{\nu\sigma}.
\end{eqnarray}
Thus, we observe that the operation of the sum (i.e.  $s_\omega + s_{\bar \omega}$)
of bosonic symmetry transportation operators $s_\omega $ and $s_{\bar \omega} $ on the {\it perfectly} anti-BRST
invariant Lagrangian density ${\cal L}_{(\bar B)} $ leads to the above transformation (32). A close look at {\it all} the individual terms on the r.h.s.
of the above equation 
establishes that all of them are expressed in terms of {\it one} of the total {\it four} CF-type restrictions on our theory. In other words, we have
the validity of the operator relationship: $s_\omega + s_{\bar \omega} = 0 $ at the level of the {\it coupled}  Lagrangian densities (as far as the symmetry considerations w.r.t. the bosonic symmetry transformation operators  $s_\omega $ and $s_{\bar \omega} $ are concerned)
provided we use {\it all} the four CF-type restrictions {\it together}. Against the backdrop of the 
above discussions, it is quite interesting to observe that the operation of the sum (i.e.  $s_\omega + s_{\bar \omega} $) of the bosonic
symmetry transformation operators on the perfectly BRST invariant Lagrangian density ${\cal L}_{(B)} $ is found to be as follows:
\begin{eqnarray}\label{33}
&& \big (s_\omega + s_{\bar \omega} \big ) \, {\cal L}_{(B)} = 
\dfrac{1}{2}\, \partial_\mu\, \Big [\big (\partial^\mu  \{\bar f^\nu + \bar F^\nu - \partial^\nu \bar C_1  \} 
- \partial^\nu \big \{\bar f^\mu + \bar F^\mu- \partial^\mu \bar C_1 \big \} \big )\, f_\nu \nonumber\\
&& + \,\big (\partial^\mu  \{ f^\nu + F^\nu - \partial^\nu C_1 \}
 - \partial^\nu  \{ f^\mu + F^\mu - \partial^\mu C_1 \} \big ) \, \bar F_\nu 
- \Big (\partial^\mu \{ B^{\nu\sigma} + \bar B^{\nu\sigma} \nonumber\\
&&- (\partial^\nu \phi^\sigma 
- \partial^\sigma \phi^\nu) \} 
+ \partial^\nu \{ B^{\sigma\mu} + \bar B^{\sigma\mu} - (\partial^\sigma \phi^\mu
- \partial^\mu \phi^\sigma) \} 
+\partial^\sigma \{ B^{\mu\nu} + \bar B^{\mu\nu} \nonumber\\
&&- (\partial^\mu \phi^\nu 
- \partial^\nu \phi^\mu) \} \Big ) \, {\cal B}_{\nu\sigma} 
+ \Big (\partial^\mu \{ {\cal B}^{\nu\sigma} + \bar{\cal B}^{\nu\sigma} - (\partial^\nu \widetilde \phi^\sigma 
- \partial^\sigma \widetilde \phi^\nu) \} + \partial^\nu \{{\cal B}^{\sigma\mu} + \bar {\cal B}^{\sigma\mu}  \nonumber\\
&&- (\partial^\sigma \widetilde \phi^\mu
- \partial^\mu \widetilde \phi^\sigma) \} +\partial^\sigma \{ {\cal B}^{\mu\nu} + \bar {\cal B}^{\mu\nu} - (\partial^\mu \widetilde \phi^\nu 
- \partial^\nu \widetilde \phi^\mu) \} \Big ) \, B_{\nu\sigma}
\Big ] 
+ \dfrac{1}{2} \;\big (\partial^\mu \bar F^\nu - \partial^\nu \bar F^\mu \big )\,\nonumber\\
&&\; \partial_\mu \big [ f_\nu +  F_\nu - \partial_\nu  C_1 \big ] 
- \dfrac{1}{2}\, \big (\partial^\mu  \{\bar f^\nu + \bar F^\nu - \partial^\nu \bar C_1  \} 
- \partial^\nu \big \{\bar f^\mu + \bar F^\mu- \partial^\mu \bar C_1 \big \} \big ) \partial_\mu  f_\nu \nonumber\\
&&+ \dfrac{1}{2} 
\big (\partial^\mu\, B^{\nu\sigma} + \partial^\nu\,B^{\sigma\mu}
+ \partial^\sigma\, B^{\mu\nu} \big ) \partial_\mu \Big [{\cal B}_{\nu\sigma} + \bar {\cal B}_{\nu\sigma} 
- (\partial_\mu \widetilde \phi_\nu 
- \partial_\nu \widetilde \phi_\mu)\Big ]  \nonumber\\
&&+ \dfrac{1}{2}\, \Big [ \partial^\mu \{ B^{\nu\sigma} + \bar B^{\nu\sigma} - (\partial^\nu \phi^\sigma 
- \partial^\sigma \phi^\nu) \} + \partial^\nu \{ B^{\sigma\mu} + \bar B^{\sigma\mu}  
- (\partial^\sigma \phi^\mu
- \partial^\mu \phi^\sigma) \} \nonumber\\
&&+ \,\partial^\sigma \{ B^{\mu\nu} + \bar B^{\mu\nu} - (\partial^\mu \phi^\nu 
- \partial^\nu \phi^\mu) \} \Big ]\, \partial_\mu {\cal B}_{\nu\sigma}.
\end{eqnarray}
A close look at the above equation demonstrates that all the {\it individual} terms on the r.h.s. incorporate into themselves {\it one} of the 
CF-type restrictions (out of the total four) that exists on our theory. This statement is true for (i) the terms that are {\it within} the square bracket
of the total spacetime derivative, and (ii)  the terms that exist {\it outside} the total spacetime derivative.
Thus, ultimately, we have been able to establish the {\it uniqueness} of the bosonic symmetry transformation operator at the level
of symmetry transformations on (i) the individual fields of our theory [cf. Eq. (26)], and (ii) the 
coupled (but equivalent) Lagrangian densities ${\cal L}_{(B)}$ 
and ${\cal L}_{(\bar B)} $ [cf. Eqs. (32),(33)]. We would like to repeat, once again, that {\it this} key observation is {\it true} only on 
the subspace of quantum fields where
{\it all} the four CF-type restrictions are satisfied {\it together} [cf. Eqs. (26),(32),(33) for more details].

It is interesting to mention that the unique
(i.e. $s_\omega = - \, s_{\bar \omega} $) bosonic symmetry transformation operator 
commutes (i.e. $[s_\omega, \; s_{(a)b} ] = 0, \; [s_\omega, \; s_{(a)d} ]  = 0$)
with {\it all} the four nilpotent symmetry transformation operators (i.e. $s_{(a)b}, \; s_{(a)d} $) of our theory. Thus, ultimately, 
we have been able to derive the following algebraic structures in the mathematical language of the continuous and discrete duality
symmetry transformation operators, namely;
\begin{eqnarray}\label{34}
&& s_{(a)b}^2 = 0, \qquad \;\; s_{(a)d}^2 = 0, \qquad \;\; s_{(a)d} = \pm\; *\; s_{(a)b} \; *,\qquad \;\;s_\omega + s_{\bar \omega} = 0, \nonumber\\
&& \{s_b, \; s_{ad} \} = 0, \qquad \{s_d, \; s_{ab} \} = 0, \qquad \{s_b, \; s_{ab} \} = 0, \quad \{s_d, \; s_{ad} \} = 0,\nonumber\\
&& \{s_b, \; s_d \} = s_\omega, \quad \{s_{ab}, \; s_{ad} \} = s_{\bar \omega}, \quad [s_\omega, \; s_{(a)b} ] = 0, \quad [s_\omega, \; s_{(a)d} ]  = 0.
\end{eqnarray}
A careful look at the above equation establishes that (ii) we have obtained the algebraic structures that are reminiscent 
of the Hodge algebra (cf. second footnote
in Sec. 1) which are satisfied by the {\it three} de Rham cohomological operators, 
and (ii) we have provided the physical realization(s) of the specific inter-relationships  that exist between the co-exterior and 
exterior derivatives (cf. Sec. 6 for more discussions). Some of the key observations, at this juncture,
are (i) the bosonic symmetry operator relationship: $s_\omega + s_{\bar \omega} = 0 $ is valid if and only if all the {\it four} CF-type restrictions are invoked 
{\it together}, (ii) the absolute anticommutativity 
properties: $\{s_b, \; s_{ab} \} = 0 $ and $\{s_d, \; s_{ad} \} = 0 $ are true only when 
the specific set of {\it three} (out of the total four) CF-type restrictions are imposed
from outside [cf. Eqs. (18),(19)], and (iii) the anticommutators $\{s_b, \; s_{ad} \} = 0 $ 
and $\{s_b, \; s_{ab} \} = 0 $ are
true only up to the U(1) vector gauge symmetry-type transformations [cf. Eq. (21)].

Before we end this section, we would like to remark on the anticommutators (20) that are found to be {\it zero} modulo the U(1) 
gauge symmetry-type transformations. To be precise,
the anticommutators $\{ s_b, \; s_{ad} \} \equiv s_\omega^{(1)}$ and 
$\{ s_d, \; s_{ab} \} \equiv s_\omega^{(2)} $ do {\it not} define a {\it new} set of 
bosonic symmetry transformations in the {\it true} sense of the word (because they raise/lower the ghost number 
by {\it two} for a  field on which 
they operate). This observation is in contradiction to the {\it fundamental} definition of a set of {\it true} bosonic symmetry 
transformations [cf. Eqs. (23),(25)]. The latter  do {\it not} change the ghost number of a field
on which they operate. To corroborate {\it this} statement, first of all, we observe the following
\begin{eqnarray}\label{35}
&&s_\omega^{(1)} \bar \beta_\mu = \partial_\mu \big (B_2 - B_3 \big ), \qquad
 s_\omega^{(1)} \phi_\mu = -\, \partial_\mu B_4,  \qquad
 s_\omega^{(1)} \widetilde \phi_\mu =  -\, \partial_\mu B_4,
 \nonumber\\
 &&
s_\omega^{(2)} \beta_\mu  = \partial_\mu \big (B_2 + B_3 \big ),\qquad
s_\omega^{(2)} \phi_\mu = +\, \partial_\mu B_5,  \qquad
s_\omega^{(2)} \widetilde \phi_\mu =  -\, \partial_\mu B_5.
\end{eqnarray}
which correspond to the transformations listed in our earlier equation (20). Under the above infinitesimal transformations, we note the following
\begin{eqnarray}\label{36}
&&s_\omega^{(1)} \, {\cal L}_{(B)} = \dfrac{1}{2}\, \partial_\mu \Big [ B_4\, \partial^\mu  \big (B_2 - B_3 \big )
- \big (B_2 - B_3 \big )\, \partial^\mu B_4 \Big ] \equiv  s_\omega^{(1)} \, {\cal L}_{(\bar B)},
\nonumber\\
&&
s_\omega^{(2)} \, {\cal L}_{(B)} = \dfrac{1}{2}\, \partial_\mu \Big [ \big (B_2 + B_3 \big )\, \partial^\mu B_5 -  B_5\, \partial^\mu  \big (B_2 + B_3 \big )
 \Big ] \equiv  s_\omega^{(2)} \, {\cal L}_{(\bar B)},
\end{eqnarray} 
which demonstrate that 
the {\it coupled}
Lagrangian densities ${\cal L}_{(B)} $ and ${\cal L}_{(\bar B)} $ transform to the total spacetime derivatives thereby rendering the 
action integral $S = \int d^4 x\, {\cal L}_{(B)} \equiv \int d^4 x\, {\cal L}_{(\bar B)} $ invariant (i.e. $s_\omega^{(1)} S = 0, \; s_\omega^{(2)} S = 0 $)
due to Gauss's divergence theorem (because of which all the physical fields vanish off as $x \to \pm \, \infty $). Thus, we conclude, ultimately, that
the continuous and infinitesimal transformations, [cf. Eq. (35)], corresponding to the bosonic operators $s_\omega^{(1)} $  and  $s_\omega^{(2)} $, are
the {\it symmetry} transformations of the action integral.

We end this very important  with the following crucial remarks. First of all, we note that the {\it true} bosonic symmetry transformation
operators $s_\omega $ and $s_{\bar \omega} $ [cf. Eqs. (23),(25)]
do {\it not} change the ghost number of the specific field of the coupled
Lagrangian densities ${\cal L}_{(B)} $ and ${\cal L}_{(\bar B)} $  when they operate on it. Second, the (anti-)ghost fields of the Lagrangian 
densities ${\cal L}_{(B)} $ and ${\cal L}_{(\bar B)} $ either do {\it not} transform at all or transform up to the U(1) gauge symmetry-type
transformations under $s_\omega $ and $s_{\bar \omega} $ [cf. Eqs. (23),(25)]. Third, we observe that the symmetry transformation operator
$s_\omega^{(1)} $ raises the ghost number of a field by {\it two} when it operates on it [cf. Eq. (35)] which implies that {\it this} transformation 
operator is {\it not} like the {\it true} bosonic symmetry transformation
operators $s_\omega $ and $s_{\bar \omega} $ [cf. Eqs. (23),(25)]. Fourth, it is interesting to point out that the symmetry operator $s_\omega^{(2)} $
lowers the ghost number by {\it two} when it operates on a field [cf. Eq. (35)]. Hence, this operator is {\it also} not like the {\it true} bosonic symmetry transformation
operators $s_\omega $ and $s_{\bar \omega} $ [cf. Eqs. (23),(25)]. Fifth, under the symmetry transformation operators $s_\omega^{(1)} $  and  $s_\omega^{(2)} $
only the bosonic vector (anti-)ghost fields $(\bar \beta_\mu)\beta_\mu $ and the (axial-)vector fields $(\widetilde \phi_\mu)\phi_\mu$ transform unlike
the transformation operators  $s_\omega $ and $s_{\bar \omega} $ [cf. Eqs. (23),(25)] under which the gauge fields as well as the fermionic and 
bosonic (anti-)ghost fields transform. Sixth, it is straightforward to note that, in the limit: $B_2 = B_3 = B_4 = B_5 = 0 $, the 
symmetry transformation operators $s_\omega^{(1)} $  and  $s_\omega^{(2)} $ disappear completely. However, the the {\it true} bosonic symmetry transformation
operators $s_\omega $ and $s_{\bar \omega} $ [cf. Eqs. (23),(25)] survive in this limit, too (along with the {\it four} 
off-shell nilpotent symmetry transformation operators [cf. Eqs. (3),(4),(9),(10)]). Finally, it is clear that if the nilpotent symmetry transformation operators
are identified with the nilpotent (co-)exterior derivatives, the Laplacian operator will be identified with (i) the {\it unique} true bosonic
symmetry transformation operator (e.g. $s_\omega$) which does {\it not} change the ghost number of the field
on which it operates, and (ii) the other (ghost number changing) bosonic symmetry transformation operators [which have been listed  in (35)] can {\it not} be, obviously, identified with the Laplacian operator of differential geometry. \\


\section{Ghost-Scale Symmetry Transformations: Algebraic Structures with Other Symmetry Operators }

The ghost parts [cf. Eqs. (2),(8)] of the {\it coupled} Lagrangian densities ${\cal L}_{(B)}$ 
and ${\cal L}_{(\bar B)} $  respect the following global (i.e. spacetime independent) scale symmetry transformations, namely;
\begin{eqnarray}\label{37}
&&C_2 \longrightarrow e^{3\,\Omega}\, C_2, \qquad \bar C_2 \longrightarrow e^{-\, 3\,\Omega}\, \bar C_2, \qquad
\beta_\mu \longrightarrow e^{2\,\Omega}\, \beta_\mu, \qquad \bar \beta_\mu \longrightarrow e^{-\, 2\,\Omega}\, \bar \beta_\mu, \nonumber\\
&&
B_4 \longrightarrow e^{2\,\Omega}\, B_4, \qquad  B_5 \longrightarrow e^{-\, 2\,\Omega}\, B_5, \qquad  C \longrightarrow e^{ \Omega}\, C, 
\qquad \bar C \longrightarrow e^{-\, \Omega}\, \bar C, \nonumber\\
&&C_1 \longrightarrow e^{\Omega}\, C_1, \qquad \bar C_1 \longrightarrow e^{-\, \Omega}\, \bar C_1,
\qquad C_{\mu\nu} \longrightarrow e^{\Omega}\, C_{\mu\nu}, \qquad \bar C_{\mu\nu} \longrightarrow e^{-\, \Omega}\, \bar C_{\mu\nu}, \nonumber\\
&&  f_\mu \longrightarrow e^{\Omega}\, f_\mu, \quad \bar f_\mu \longrightarrow e^{-\, \Omega}\, \bar f_\mu, \quad
F_\mu \longrightarrow e^{\Omega}\, F_\mu, \;\; \bar F_\mu \longrightarrow e^{ -\,\Omega}\, \bar F_\mu, 
\;\; \Psi \longrightarrow e^0\,  \Psi,
\end{eqnarray}
where (i) the factor $\Omega $ is a {\it bosonic} spacetime independent (i.e. global) scale parameter,
 (ii) the numerical factors, in the exponents, denote the ghost numbers of the (anti-)ghost fields, and (iii)  the generic field
$\Psi
$ stands for {\it all} the basic and auxiliary fields that are present in the non-ghost parts [cf. Eqs. (1),(7)]
of the above {\it coupled} Lagrangian densities and they
carry the ghost number equal to {\it zero}. For the sake of brevity, we set the global scale parameter equal to one (i.e. $\Omega = 1 $).
As a consequence of this choice, we obtain the continuous and infinitesimal version of the above ghost-scale symmetry transformations as
\begin{eqnarray}\label{38}
&&s_g C_2 = +\, 3\, C_2, \qquad s_g \bar C_2 = -\, 3\, \bar C_2, \qquad
s_g \beta_\mu = +\, 2\,\beta_\mu, \qquad s_g \bar \beta_\mu = -\, 2\, \bar \beta_\mu, \nonumber\\
&&
s_g B_4 = +\,2\, B_4, \qquad  s_g B_5 = -\, 2\, B_5, \qquad  s_g C = +\, C, 
\qquad s_g \bar C = -\,  \bar C, \nonumber\\
&& s_g C_1 = +\, C_1, \qquad s_g \bar C_1 = -\, \bar C_1,
\qquad s_g C_{\mu\nu} = +\, C_{\mu\nu}, \qquad s_g \bar C_{\mu\nu} = -\,  \bar C_{\mu\nu}, \nonumber\\
&&  s_g f_\mu = +\, f_\mu, \quad s_g \bar f_\mu = -\,  \bar f_\mu, \quad
s_g F_\mu = +\, F_\mu, \quad s_g \bar F_\mu = -\, \bar F_\mu,  \quad
s_g \Psi = 0,
\end{eqnarray}
where we have denoted the continuous and infinitesimal version of the ghost-scale symmetry transformation by the symbol $s_g$. It is 
straightforward to note that, under {\it this} infinitesimal ghost-scale symmetry transformations, we obtain: $s_g {\cal L}_{(B)} = 0, \;
s_g {\cal L}_{(\bar B)} = 0 $ which establishes the fact that the above continuous and infinitesimal ghost-scale  transformations
(37) are the {\it perfect} symmetry transformations for the {\it coupled} Lagrangian densities ${\cal L}_{(B)}$ 
and ${\cal L}_{(\bar B)} $.

At this stage of our discussions, as far as the {\it continuous} and infinitesimal symmetry operators of our theory are concerned, 
we have (i) a set of {\it four} nilpotent (anti-)BRST and (anti-)co-BRST symmetry
transformation operators, (ii) a couple of bosonic symmetry transformation operators [cf. Eq. (21)] in our theory which satisfy
the relationship: $s_\omega + s_{\bar \omega} = 0 $ on the subspace of the quantum fields (when we invoke 
the validity of {\it all} the four CF-type restrictions 
that are present on our theory), (iii) a couple of {\it additional} bosonic symmetry transformation operators [cf. Eq. (20)], and (iv)
the infinitesimal ghost-scale symmetry transformations (38). It will be interesting to obtain the algebraic structure between
the infinitesimal ghost scale symmetry transformation operator $s_g$ and the {\it rest} of the continuous symmetry operators of our theory.
In this context, we observe the following
\begin{eqnarray}\label{39}
&& \big [s_g, \; s_b \big ] = +\, s_b,  \qquad \big [s_g, \; s_{ab} \big ] = -\, s_{ab}, \qquad \big [s_g, \; s_d \big ] = -\, s_d, \nonumber\\
&& \big [s_g, \; s_{ad} \big ] = +\, s_{ad},  \qquad \big [s_g, \; s_{\omega} \big ] = 0, \qquad \big [s_g, \; s_{\bar \omega} \big ] = 0,
\end{eqnarray}
which establish that the ghost number of a field is raised by one when it is operated upon by the BRST symmetry transformation operator $s_b$
as well as the anti-co-BRST symmetry transformation operator $s_{ad}$ [cf. Eqs. (4),(9)]. On the other hand, the operation of the co-BRST symmetry
transformation operator $s_d$ as well as the anti-BRST symmetry transformation operator $s_{ab}$ 
leads to the lowering the ghost number of a  {\it specific} field by one [cf. Eqs. (3),(10)]. A close look at the bosonic symmetry transformations
[cf. Eqs. (23),(25)]
(that are generated by the operators $s_\omega$ and $s_{\bar\omega}$) establish that the ghost number of a field remains {\it intact} when it is acted
upon by the {\it true} set [cf. Eq. (21)]
of the bosonic symmetry transformation operators (i.e. $s_\omega = \{s_b, \; s_d \} $ and $s_{\bar\omega} = \{s_{ab}, \; s_{ad} \} $).

We end this section with the following decisive remarks. First of all, we note that the ghost number considerations enable us to cluster together
a pair of infinitesimal and continuous symmetry transformation operators as: $(s_b, \; s_{ad})$, $(s_d, \; s_{ab})$ and $(s_\omega, \; s_{\bar\omega})$.
Second, we note that the {\it true} bosonic symmetry transformation operators commute with {\it all} the continuous and infinitesimal symmetry
transformation operators of our theory {\it including} the ghost-scale symmetry transformation operator [cf. Eq. (39)]. Finally, the 
infinitesimal and continuous ghost-scale symmetry transformation operator $s_g$ distinguishes (and differentiates) between the {\it true}
bosonic symmetry transformation operators (e.g. $s_\omega$ and $s_{\bar\omega}$) and the {\it fictitious} bosonic symmetry
transformation operators (e.g. $s_\omega^{(1)}$ and $s_{\bar\omega}^{(2)}$).  To be precise, we
observe the following algebraic structure between the ghost-scale symmetry transformation operator $s_g$ and the {\it fictitious} set of 
(ghost number changing) bosonic symmetry
transformation operators (i.e. $s_\omega^{(1)}$ and $s_{\bar\omega}^{(2)}$), namely; 
\begin{eqnarray}\label{40}
\big [s_g, \; s_{\omega}^{(1)} \big ] = +\, 2\,  s_{\omega}^{(1)}, \qquad \big [s_g, \; s_{\omega}^{(2)} \big ] = -\, 2\,  s_{\omega}^{(2)}.
\end{eqnarray} 
In other words, we note the the {\it fictitious} set of bosonic symmetry
transformation operators (i.e. $s_\omega^{(1)}$ and $s_{\bar\omega}^{(2)}$) do {\it not} commute with the infinitesimal and continuous
ghost-scale symmetry transformation operator $s_g$. The algebraic structure (40) demonstrates that the fictitious bosonic symmetry
transformation operators $s_\omega^{(1)}$ and $s_{\bar\omega}^{(2)}$ raise/lower the ghost number of a field by a factor of two when they operate 
on this {\it specific} field [cf. Eq. (35)]. Hence, the bosonic symmetry 
transformation operators  $s_\omega^{(1)}$ and $s_{\bar\omega}^{(2)}$ are
{\it not} interesting to us in view of our main objective to establish that our present 4D BRST-quantized field-theoretic system is 
a tractable example for Hodge theory. We call the (ghost number {\it changing}) bosonic symmetry transformation operators
$s^{(1)}_\omega $ and $s^{(2)}_\omega $ as the ``fictitious'' transformation operators 
because they do {\it not} behave like the Laplacian operator of differential geometry.
At the moment, we are {\it not} aware of a set of {\it two} geometrical operators (within the realm of differential geometry) that raise/lower the
degree of a given form by $\pm \, 2 $.  \\


\section{Summary and Future Perspective}

In our present endeavor, we have very briefly sketched the existence of the infinitesimal, continuous 
and off-shell nilpotent (i.e. $s_{(a)b}^2 = 0, \;s_{(a)d}^2 = 0 $)
versions of the (anti-)BRST and (anti-)co-BRST symmetry transformation operators (i.e. $s_{(a)b} $ and $s_{(a)d} $, respectively) for the coupled
Lagrangian densities ${\cal L}_{(B)} $ and ${\cal L}_{(\bar B)} $ that describe the 4D BRST-quantized {\it combined} field-theoretic system of the free
Abelian 3-form and 1-form gauge theories. We would like to lay emphasis on the fact that {\it this} combined field-theoretic system is essential
for the {\it existence} of the nilpotent (anti-)co-BRST symmetry transformations. As far as the nilpotent (anti-)BRST symmetry transformations are 
concerned, we note that {\it these} transformations are respected by the BRST-quantized versions of the free Abelian 3-form and 1-form gauge
theories separately (and independently) in any arbitrary D-dimension of spacetime. 
We have given a great deal of importance to the existence of a couple of discrete duality symmetry
transformations (in the context of our 4D BRST-quantized theory) and have highlighted their roles in the algebraic structures that are obeyed by the {\it four} 
off-shell nilpotent, infinitesimal and continuous symmetry transformation 
operators and the inter-relationships that exist amongst them [cf. Eqs. (14),(17),(34)]. To be precise, we have provided the 
physical realizations [cf. Eqs. (14),(17)] of the well-known mathematical relationship $\delta = \pm\, *\, d\, * $ between 
the nilpotent (i.e. $ d^2 = 0, \delta^2 = 0$)
(co-)exterior derivatives $(\delta)d$ of differential geometry in terms of the interplay between (i) the discrete duality symmetry transformation operators
[cf. Eqs. (13),(16)], and (ii) the infinitesimal, continuous and off-shell  nilpotent symmetry transformation 
operators [cf. Eqs. (3),(4),(9),(10)].

Against the backdrop of the above paragraph,
 we would like to emphasize that the relationships in equations (14) and (17) corroborate  our observations that the nilpotent (i.e. $s_{(a)d}^2 = 0, \; s_{(a)b}^2 = 0 $) (anti-)co-BRST
symmetry transformations $s_{(a)d} $ [cf. Eqs. (9),(3)] and (anti-)BRST symmetry transformations $s_{(a)b} $ [cf. Eqs. (10),(4)]
are {\it dual} to one-another and the precise choices of the ($\pm$) signs 
in their relationships [cf. Eqs. (14),(17)] are dictated by the set rules [cf. Eq. (15)]
of {\it duality-invariant} Abelian $p$-form ($ p = 1, 2, 3 ...$) gauge theories
that have been propounded in [19]. One of the highlights of our present investigation is the observation that the {\it reciprocal} 
relationships: $s_{(a)b} = \pm\; *\; s_{(a)d} \; * $ {\it also} exist in our 4D BRST-quantized theory. However, the choices of the  ($\pm$) signs
are {\it not} dictated by any kinds of set rules. For the sanctity and the proof of {\it these} relationships (i.e. $s_{(a)b} = \pm\; *\; s_{(a)d} \; * $),
one has to make the choices for the ($\pm$) signs judiciously. A close look at the nilpotent transformations [cf. Eqs. (3),(4),(9),(10)] 
demonstrate that the pair of nilpotent operators ($s_b, \; s_{ad} $) raise the ghost number of a field by one. On the other hand,
the nilpotent pair of operators ($s_d, \; s_{ab} $) lower the ghost number of a field by one (cf. Sec. 5 for more comments). 
These observations are just like the operations
of the exterior and co-exterior derivatives of differential geometry on a given form. Thus, 
we find that (i) the algebraic structures of the symmetry 
transformation operators (cf. Secs. 3-5), and (ii) the ghost number considerations (in the context of the
off-shell nilpotent symmetry operators)
are such that we obtain a two-to-one mapping [i.e. $(s_b, \; s_{ad}) \Rightarrow d, \; (s_d, \; s_{ab}) \Rightarrow \delta,
\; (s_\omega, \; s_{\bar \omega}) \Rightarrow \Delta$] between (i) the symmetry transformation operators, and (ii) the 
de Rham cohomological operators
(where the discrete duality symmetry operators [cf. Eqs. (13),(16)] provide the physical realization(s) of the Hodge duality $*$ operator in the
mathematical relationship: $\delta = \pm\, *\, d\, * $ between the nilpotent (i.e. $d^2 = 0, \, \delta^2 = 0 $)
(co-)exterior derivatives $(\delta)d$).

Out of the total {\it six} anticommutators (that can be constructed from the {\it four} nilpotent symmetry operators), we have shown that (i) the {\it two}
specific anticommutators (i.e. $\{ s_d, \; s_{ab} \} = 0 $ and $\{ s_{ad}, \; s_{b} \} = 0 $) are {\it trivially} zero on their own
for {\it all} the quantum fields of our 4D  BRST-quantized theory {\it except} for the fields $\beta_\mu, \; \bar \beta_\mu, \; \phi_\mu$ and
$\widetilde \phi_\mu $ 
which transform up to the U(1) {\it vector} gauge symmetry-type transformations [cf. Eq. (35)], (ii) another
{\it two} anticommutators  (i.e. $\{ s_d, \; s_{ad} \} = 0 $ and $\{ s_{ab}, \; s_{b} \} = 0 $) are {\it zero} provided we use the 
specific set of {\it three} (out of the total {\it four}) CF-type restrictions on our theory, and (iii) the rest of the two anticommutators
(i.e. $\{ s_d, \; s_{b} \} = s_\omega $ and $\{ s_{ab}, \; s_{ad} \} = s_{\bar \omega}$) define a set of two {\it bosonic} symmetry
transformation operators $s_\omega $  and $s_{\bar \omega} $. We have established, in our present investigation, that there exists a {\it unique} bosonic
symmetry operator (i.e. $s_\omega  = -\,s_{\bar \omega}   $) on the {\it subspace} of 
the quantum fields which is precisely defined by {\it all} the four CF-type restrictions (i.e. $B_{\mu\nu} + \bar B_{\mu\nu} 
= (\partial_\mu  \phi_\nu  - \partial_\nu  \phi_\mu),\; {\cal B}_{\mu\nu} + \bar {\cal B}_{\mu\nu} 
= (\partial_\mu \widetilde \phi_\nu  - \partial_\nu \widetilde \phi_\mu), \; f_\mu + F_\mu = \partial_\mu C_1, \;
\bar f_\mu + \bar F_\mu = \partial_\mu \bar C_1 $) that exist on our 4D BRST-quantized field-theoretic system  [cf. Eqs. (28),(30)--(33)].
It is worthwhile to point out that the anticommutators  $\{ s_b, \; s_{ad} \} $ and $\{ s_{ab}, \; s_{d} \}$
do {\it not} define 
a new set of {\it true} bosonic symmetry transformations because they (i) change the ghost numbers of the specific set of
fields $\beta_\mu, \; \bar \beta_\mu, \; \phi_\mu$ and $\widetilde \phi_\mu $ of our  4D BRST-quantized theory by $\pm\, 2 $  [cf. Eq. (35)], 
(ii) transform  {\it these} specific fields {\it only} up to the U(1) {\it vector} gauge symmetry-type transformations, and (iii) disappear 
in the limit: $B_2 = B_3 = B_4 = B_5 = 0$
(which is {\it not} the case with the {\it true} bosonic  symmetry transformations $s_\omega $ and $s_{\bar \omega} $ 
that do {\it not} change the ghost number of any arbitrary field of our theory when they operate on it [cf. Eqs. (23),(25)]). Thus, for 
all practical purposes, the ghost number {\it changing} bosonic symmetry transformation operators $s_\omega^{(1)} $ and  $s_\omega^{(2)} $
are taken to be zero modulo the U(1) {\it vector} gauge symmetry-type transformations [cf. Eq. (35)].

The study 
of the BRST-quantized field-theoretic systems, as the tractable examples
for Hodge theory, has been mathematically and physically useful in the sense that 
we have been able to establish that the 2D (non-)Abelian gauge theories (without any interaction with the matter fields) provide a {\it new}
type of topological field theory (TFT) that captures [20] a few aspects of the Witten-type TFTs [21] and some salient features of the 
Schwarz-type TFTs [22]. In addition,
such studies have led to the existence of ``exotic'' fields with the {\it negative} kinetic terms (see, e.g. [18],[23],[24] and references therein)
that are a set of possible candidates for (i) the ``phantom''  and/or ``ghost'' fields
of the cyclic, bouncing and self-accelerated cosmological models of the Universe (see, e.g. [25-27]and references therein), 
and (ii) the dark energy/dark matter (see, e.g. [28,29] and references therein).

In our future endeavor, we plan to derive (i) all the Noether conserved charges corresponding to {\it all} the continuous 
and infinitesimal symmetry transformations 
of our 4D BRST-quantized theory, and (ii) the algebraic structures satisfied by the {\it appropriate} conserved charges and study
the mapping(s) between the {\it appropriate} conserved charges and the de Rham cosmological operators of differential geometry. 
A very interesting future direction is to study the 4D BRST-quantized field-theoretic system  of the St{\" u}ckelberg-modified  {\it massive} version
of our {\it present} theory and discuss its discrete and continuous symmetry transformations (and corresponding conserved charges).
In other words, we
envisage to prove that (i) our present 4D BRST-quantized field-theoretic system, and
(ii) the St{\" u}ckelberg-modified {\it massive} version of our present {\it combined}
field-theoretic system of the free Abelian 1-form and 3-form gauge theories, are a set of  examples for Hodge theory. \\

\vskip0.8cm

\noindent
{\bf {\Large Acknowledgments}}\\

\noindent
Fruitful conversations with R. Kumar, on the subject matter of our present endeavor, are gratefully acknowledged. Thanks  are also due to
our esteemed Reviewer for very useful comments which have made our presentation more transparent and readable.\\

\vskip 1cm

\noindent
{\bf Declarations}\\

\noindent
{\bf Funding statement}\\

\noindent
No funding was received for this research.\\

\noindent
{\bf Declaration of competing interest}\\

\noindent
The author declares that he has no known competing financial interests or personal relationships that could have appeared
to influence the work reported in this paper.\\

\noindent
{\bf Data availability}\\

\noindent
No data was used for the research described in the article.\\


\vskip 1.3cm
\begin{center}
{\bf Appendix A: On the Glossary of Fields}
\end{center}

\noindent
In this Appendix, we provide the list of {\it all} the fields that are present in the coupled (but equivalent) BRST and anti-BRST invariant Lagrangian densities ${\cal L}_{(B)} = {\cal L}_{(NG)} + {\cal L}_{(FP)} $ [cf. Eqs. (1)(2)] and 
${\cal L}_{\bar B} = {\cal L}_{(ng)} + {\cal L}_{(fp)} $ [cf. Eqs. (7),(8)], respectively . We lay emphasis on their bosonic/fermionic, basic/auxiliary, independent/restricted, tensor/vector/scalar, etc., nature and mention clearly their ghost number(s), too. Since there is presence of a tower of
different kinds (as well as varieties) of fields in 
the {\it above} coupled Lagrangian densities of our theory, for the sake of readers' convenience, 
we collect {\it all} these fields (of our 4D BRST-quantized field-theoretic system) in the following tabulated form:

\begin{table}[h!]
\centering
\begin{tabular}{ |p{1.2cm}|p{1.9cm}|p{1.99cm}|p{2.5cm}|p{1.2cm}|p{4.2cm}| }
\hline
 \textbf{Fields} & \textbf{Bosonic/ Fermionic} & \textbf{Basic/Au- xiliary}
 & \textbf{Tensorial Nature} & \textbf{Ghost Number} & \textbf{Independent/Restr- icted} \\
 \hline
 $A_{\mu\nu\lambda}$  & Bosonic  & Basic & Totally Antisymmetric & 0 & Independent  \\
 $A_\mu$  & Bosonic  & Basic & Vector & 0 & Independent  \\
 $C_{\mu\nu} $ & Fermionic & Basic & Antisymmetric & + 1 &  Independent \\
 $\bar C_{\mu\nu} $  & Fermionic & Basic & Antisymmetric & - 1 &  Independent\\
 $B_{\mu\nu}$ & Bosonic & Auxiliary & Antisymmetric & 0 & Restricted   \\
 $\bar B_{\mu\nu}$  & Bosonic & Auxiliary & Antisymmetric & 0 & Restricted    \\
 ${\cal B}_{\mu\nu}$ & Bosonic & Auxiliary & Antisymmetric & 0 & Restricted   \\
 $\bar {\cal B}_{\mu\nu}$  & Bosonic & Auxiliary & Antisymmetric & 0 & Restricted    \\
 $\beta_\mu$ & Bosonic & Basic & Vector & + 2 & Independent \\
 $\bar \beta_\mu$ & Bosonic & Basic & Vector & - 2 & Independent \\
 $C$  & Fermionic  &  Basic & Scalar & +1 & Independent \\
$\bar C$  & Fermionic  &  Basic & Scalar & - 1 & Independent  \\
 $C_1$  & Fermionic  &  Basic & Scalar & + 1 & Restricted  \\
$\bar C_1$  & Fermionic  &  Basic & Scalar & - 1 & Restricted   \\
 $C_2$  & Fermionic  &  Basic & Scalar & + 3 & Independent  \\
 $\bar C_2$  & Fermionic  &  Basic & Scalar & - 3 & Independent  \\
 $\phi_\mu$ & Bosonic  &  Basic & Vector & 0 & Restricted \\
 $\widetilde \phi_\mu$ & Bosonic  &  Basic & Axial-vector & 0 & Restricted \\
$B$ & Bosonic  &  Auxiliary & Scalar & 0 & Independent  \\
 $B_1$  & Bosonic  &  Auxiliary & Pseudo-scalar & 0 & Independent  \\
 $B_2$  & Bosonic  &  Auxiliary & Scalar & 0 & Independent  \\ 
 $B_3$  & Bosonic  &  Auxiliary & Pseudo-scalar & 0 & Independent  \\
 $B_4$  & Bosonic  &  Auxiliary & Scalar & + 2 & Independent  \\
  $B_5$  & Bosonic  &  Auxiliary & Scalar & - 2 & Independent  \\
 $f_\mu$ & Fermionic & Auxiliary & Vector & + 1 & Restricted  \\
 $\bar f_\mu$  &  Fermionic & Auxiliary & Vector & - 1 & Restricted  \\
 $F_\mu$ &  Fermionic & Auxiliary & Vector & + 1 & Restricted  \\
 $\bar F_\mu$ & Fermionic & Auxiliary & Vector & - 1 & Restricted  \\
 \hline 
\end{tabular}

\vskip 0.5cm

\caption{Tower of fields and their specifications}
\end{table}

\noindent
It is worthwhile to mention that, in the above table, the word ``restricted'' (against a specific field) denotes that
{\it this} field is a participant in one of the {\it four} (anti-)BRST and (anti-)co-BRST invariant CF-type restrictions
that exist on our theory.\\

\end{document}